\documentclass[10pt]{iopart}

\newcommand{\wh}[1]{\widehat{#1}}
\newcommand{\wt}[1]{\widetilde{#1}}
\newcommand{\eps}{\varepsilon}

\newcommand{\Ttm}[1]{\langle {#1} \rangle} 

\newcommand{\rms}[1]{{#1}_{\rm rms}} 

\newcommand{\avg}[1]{\ensuremath{\langle #1 \rangle}}

\newcommand{\Sn}{\Phi} 
\newcommand{\snw}{\varphi} 
\newcommand{\taud}{\ensuremath{\tau_\mathrm{d}}}

\newcommand{\Snnorm}{\ensuremath{\widetilde{\Phi}}} 

\newcommand{\nee}{\ensuremath{n_\mathrm{e}}}
\newcommand{\Te}{\ensuremath{T_\mathrm{e}}}
\newcommand{\isat}{\ensuremath{I_\mathrm{sat}}}
\newcommand{\Isat}{\ensuremath{\widetilde{I}_\mathrm{sat}}}
\newcommand{\Vf}{\ensuremath{V_\mathrm{f}}}
\newcommand{\nebar}{\ensuremath{\overline{n}_\mathrm{e}}}
\newcommand{\ngw}{\ensuremath{n_\mathrm{G}}}
\newcommand{\Ip}{\ensuremath{I_\mathrm{P}}}
\newcommand{\Bt}{\ensuremath{B_\mathrm{T}}}

\newcommand{\neng}{\ensuremath{\overline{n}_\mathrm{e} / n_\mathrm{G}}}
\newcommand{\fg}{\ensuremath{f_\mathrm{GW}}}

\usepackage[pdftex]{graphicx}
\usepackage{iopams} 
\usepackage{bm}
\usepackage{fancybox}
\usepackage[dvipsnames]{xcolor}
\usepackage{multirow}
\usepackage[section]{placeins}
\usepackage{amsmath}
\usepackage{subfigure}
\usepackage[colorlinks]{hyperref}
\usepackage{cite}
\usepackage[toc,page]{appendix}
\usepackage{doi}
\hypersetup{
    colorlinks=true,
    linkcolor=blue,
    citecolor=blue,
    urlcolor = blue,
}

\usepackage{tikz}
\usetikzlibrary{matrix}
\usetikzlibrary{shapes.geometric, arrows, shadows}
\usetikzlibrary{fit,backgrounds} 

\begin{document}

\title[Intermittent far scrape-off layer fluctuations in Alcator C-Mod]{Strongly intermittent far scrape-off layer fluctuations in Alcator C-Mod plasmas close to the empirical discharge density limit}

\author{Sajidah Ahmed,$^{1}$ Odd Erik Garcia,$^{1}$ Adam Q Kuang$^{2}$, Brian LaBombard,$^{3}$ James L Terry$^{3}$ and Audun Theodorsen$^{1}$}

\address{$^{1}$
Department of Physics and Technology, UiT The Arctic University of Norway, N-9037 Tromsø, Norway
}
\address{$^{2}$
Commonwealth Fusion Systems, 117 Hospital Road, Devens, MA 01434, United States of America
}
\address{$^{3}$
MIT Plasma Science and Fusion Center, Cambridge, MA 02139, United States of
America
}
\ead{sajidah.ahmed@uit.com}
\vspace{10pt}
\begin{indented}
\item[]\today
\end{indented}

\begin{abstract}
Intermittent plasma fluctuations in the boundary region of the Alcator C-Mod device were comprehensively investigated using data time-series from gas puff imaging and mirror Langmuir probe diagnostics. Fluctuations were sampled during stationary plasma conditions in ohmically heated, lower single null diverted configurations with scans in both line-averaged density and plasma current, with Greenwald density fractions up to $0.85$. Utilizing a stochastic model, we describe the plasma fluctuations as a super-position of uncorrelated pulses, with large-amplitude events corresponding to blob-like filaments moving through the scrape-off layer. A deconvolution method is used to estimate the pulse arrival times and amplitudes. The analysis reveals a significant increase of pulse amplitudes and waiting times as the line-averaged density approaches the empirical discharge density limit. Broadened and flattened average radial profiles are thus accompanied by strongly intermittent and large-amplitude fluctuations. Although these filaments are arriving less frequently at high line-averaged densities, we show that there are significant increases in radial far-SOL particle and heat fluxes which will further enhance plasma--wall interactions. The stochastic model has been used as a framework for study of the scalings in the intermittency parameter, flux and mean amplitude and waiting times, and is being used to inform predictive capability for the effects of filamentary transport as a function of Greenwald fraction.
\end{abstract}


\submitto{\PPCF}
\maketitle

\section{Introduction}\label{sec:introduction}

The boundary of magnetically confined fusion plasmas plays an imperative role in determining the heat flux density onto the material surfaces. The open field line region, known as the scrape-off layer (SOL), is riddled with turbulent flows mainly in the form of blob-like filaments that propagate radially toward the vessel walls \cite{labombard-2001,pitts-2005,dippolito-2012,Birkenmeier2015}. These are believed to be the dominant contributors to the cross-field transport of particles and heat in the SOL, which can lead to enhanced erosion rates of the wall materials and threaten the quality of plasma confinement \cite{stangeby-book,krasheninnikov2020edge,pitts-2005,lipschultz-2007,Marandet2011}. For reactor-relevant devices, reliable predictions of the expected plasma--wall interactions are necessary to mitigate the deleterious effects on the wall and the inevitable sputtering of material atoms. Therefore, the statistical properties of the plasma fluctuations are of great interest in order to predict and monitor the far-SOL transport, in particular the amplitudes of these far-SOL plasma fluctuations, their frequency of occurence and the duration times. 

Filaments are coherent pressure perturbations with order-unity relative fluctuation amplitudes, elongated along and localized perpendicular to the magnetic field lines \cite{zweben-1985,grulke-2006,zweben-2002,terry-2003,zweben-2004,Terry2005,maqueda-2011,banerjee-2012}. They are believed to originate in the vicinity of the last closed magnetic flux surface (LCFS), and are observed in all magnetic configurations and confinement states \cite{endler-1999, labombard-2000, boedo-2001, antar-2001-prl,terry-2003,zweben-2007,xu-2009,cziegler-2010,maqueda-2011}. On the low-field side of the SOL, magnetic gradient and curvature drifts lead to electric polarization of these filaments, resulting in a radial propagation towards the vessel walls \cite{krash-2001,bian-2003,bisai-2005,garcia-2006,garcia-2006-ps}.

In single-point measurements of SOL plasma fluctuations, filaments are observed as large-amplitude bursts, and have been shown to exhibit several robust statistical properties \cite{boedo-2001,antar-2001, rudakov-2002,boedo-2003,graves-2005,rudakov-2005,garcia-2007-coll,garcia-2007-nf,garcia-2015,theodorsen-2016-ppcf,garcia-2013,garcia-2013-jnm}. These include skewed and flattenened probability density functions (PDFs) with elevated tails for large amplitudes and frequency spectra which are flat for low frequecies and power-law like for high frequencies \cite{terry-2003, antar-2003,horacek-2005,Agostini2007nstx,walkden-2017,theodorsen-2017-nf,theodorsen-2018-php,garcia-2018,kube-2018,kube-2020,kuang-2019,zurita2022stochastic}. Conditional averaging revealed exponentially distributed pulse amplitudes and waiting times as well as the waveform of the bursts to be well described by a two-sided exponential function \cite{rudakov-2002,boedo-2003,garcia-2007-nf,garcia-2007-coll,banerjee-2012,carralero-2014,theodorsen-2016-ppcf,garcia-2017-nme,garcia-2018,zurita2022stochastic}.

To describe the statistical properties of the SOL fluctuations, a stochastic model based on a super-position of uncorrelated exponential pulses with constant duration has been introduced \cite{garcia-2012,garcia-2016,militello-2016,garcia-2017-ac,theodorsen-2017-ps,theodorsen-2018-ppcf}. It can be shown that exponentially distributed amplitudes and waiting times between consecutive events lead to frequency power spectral densities (PSDs) with a Lorentzian shape and Gamma probability density functions (PDFs) \cite{garcia-2012,garcia-2016,garcia-2017-ac,garcia-2017-lorentz,theodorsen-2017-ps,theodorsen-2017-nf,theodorsen-2018-ppcf}. Both of these properties have been shown to be in excellent agreement with Langmuir probe and gas puff imaging (GPI) measurements in various confinement regimes and fusion devices \cite{theodorsen-2016-ppcf,theodorsen-2017-nf,garcia-2017-nme,theodorsen-2018-php,garcia-2018}. Operating at a sampling rate of a few megahertz, these diagnostics have sampling times relevant for studying SOL turbulence dynamics. Further, the stochastic model has been used to validate numerical simulations of SOL turbulence, which reveal the same statistical properties of the fluctuations as found from experimental measurements \cite{decristoforo2021numerical}.

The stochastic model describes the fluctuations at any given position in the SOL as only due to blob-like filaments. This notion probably does not apply in the vicinity of the LCFS and near-SOL region due to the presence of drift waves and a shear layer where poloidal velocities may be significant \cite{terry-2001,labombard2003toroidal,labombard-2005,Terry2005,zweben-2010,agostini-2011,Russell2016}. However, the focus of this investigation is large-amplitude fluctuations in the far-SOL where we expect low poloidal velocities and radial motion of the blob structures to dominate \cite{terry-2001,Terry2005,zweben-2010,agostini-2011,kube-2013,offeddu_2022}. Moreover, the distinction between the near and the far-SOL regions disappears at high line-averaged densities, supporting the assumption that the most dominant process leading to these flat and broad profiles are the filaments \cite{rudakov-2001,labombard2002interpretation,Whyte2005,asakura-2007,vianello-2010,dippolito-2011,carralero-2014,mccormick-1992,Militello2016,walkden-2017}. Measurements from various fusion devices suggest that edge and SOL transport appears to be strongly related to the empirical discharge density limit, as seen by the broadening and flattening of the radial SOL density and temperature profiles \cite{labombard-2001,boedo-2001,labombard-2005,Antar2005scaling,garcia-2007-nf,garcia-2007-coll,guzdar2007large}. As one approaches the density limit, these filamentary structures are observed on closed-field line regions as a result of the far-SOL extending all the way inside the LCFS \cite{terry-2003}.

In this contribution, exceptionally long fluctuation time series from the GPI and the mirror Langmuir probe (MLP) on Alcator C-Mod were analyzed in order to carry out a study of the stochastic model parameters with the line-averaged density of the main plasma and the plasma current. From the measurement data, the parameters of the stochastic model are estimated, comprising the mean pulse amplitude and the average pulse duration and waiting times. By reformulating the stochastic model as a convolution of the pulse function with a train of delta pulses, the Richardson--Lucy (RL) deconvolution algorithm was used to recover the pulse amplitudes and their arrival times from the measurement time series \cite{richardson-1972,lucy-1974,benvenuto-2010}. This provides an accurate determination of these quantities for both small and large-amplitude events \cite{theodorsen-2018-php,ahmed-2022}, as opposed to the much-used conditional averaging technique that selectively measures properties only of the large-amplitude events. Concerns about using the conditional averaging technique arise from choice of amplitude threshold and the struggle to deal with significant pulse overlap and noise consistently. The application of the deconvolution method was initially used to analyze GPI data from Alcator C-Mod \cite{theodorsen-2018-php}. Subsequently, this method underwent theoretical investigations in reference \cite{ahmed-2022}, resulting in significant improvements. In our present study, we employ this enhanced version of the deconvolution method for the first time, showcasing its effectiveness in our analyses. Further, in previous studies using the deconvolution algorithm, the method contained a free parameter chosen to make the mean estimated waiting time from the deconvolution as close as possible to the mean waiting time estimated from $\gamma / \taud$ \cite{theodorsen-2018-php, kube-2020}. This application of the deconvolution method on the measurement time-series in this study does not employ this free parameter.

Estimating various blob velocity scaling regimes and connection to the divertor target is beyond the scope of this study. This is because there is no consensus in the literature about the role that the divertor conditions play in the main plasma SOL. References that address this issue, but with conflicting conclusions, are \cite{carralero-2014,carralero-2015,carralero-2017, wynn-2018,kuang-2019}. Here we restrict the study to the main plasma SOL alone since existing work on Alcator C-Mod in reference \cite{kuang-2019} has shown no obvious effect of the divertor state on the SOL. Blob regimes have previously been studied in Alcator C-Mod, assessing the role of the Greenwald density fraction \cite{kube-2013,kube-2016-php,kuang-2019}. It was found that local plasma parameters appeared to have a strong influence on radial propagation, which is not captured well by blob theory. Furthermore, the results of the tracking of blob-structures from reference \cite{kube-2013} suggest that the blob size changes little with line-averaged density.

Here, results are presented from detailed time series analyses of far-SOL fluctuation data in ohmically heated, lower single null (LSN) diverted Alcator C-Mod plasmas with a wide range of line-averaged densities ($0.46\times10^{20}-2.8\times10^{20}\,\text{m}^{-3}$) and plasma currents ($0.53-1.1\,\text{MA}$). These line-averaged densities and plasma currents will be quoted in terms of the Greenwald fraction, $\fg = \neng$, where $\nebar$ is the line-averaged density and the empirical Greenwald density limit is defined as $\ngw = (\Ip/\pi a^2) \times 10^{20}$ m$^{-3}$, where $a$ is the minor radius of the plasma in units of meters and $\Ip$ is the plasma current in units of mega-Amperes \cite{greenwald-2002}. The study by reference \cite{labombard-2001} demonstrated that as the empirical density limit was approached in Alcator C-Mod, there was a noticeable increase in the radial particle and heat fluxes within the scrape-off layer (SOL). This finding serves as additional motivation for utilizing the Greenwald density fraction as a parameter of interest. Incorporating the Greenwald fraction enables comparisons between different machines and facilitates predictive capabilities, both of which are key benefits of this work. In this study, we present new findings that focus on the exploration of Greenwald fractions within the range of $0.48<\fg \leq 0.85$ in the framework of stochastic modeling. Previous studies \cite{garcia-2013,theodorsen-2017-nf,kuang2019measurements,kube-2013} have not investigated this specific range. We provide explicit evidence illustrating the changes in fluctuation statistics across a wide range of Greenwald fractions. These findings are subsequently employed to interpret flux measurements in the far-SOL region. Although a discrepancy has been indicated in reference \cite{kube-2020} between the GPI and the MLP, we emphasize the discrepancies between these two diagnostics across a wide density scan range, showing that distinct trends exist with the fluctuation statistics. We show that it is the functional dependence itself which is different, not just the parameter values. Despite these differences, both diagnostics show that strongly intermittent fluctuations are observed as the density limit is approached.

This contribution is structured in the following way: Section \ref{sec:experimental-setup} gives details of the experimental setup and a brief overview of the GPI and MLP diagnostic systems. In section \ref{sec:stat-framework}, we review the stochastic model and the base case (that is, exponentially distributed pulse amplitudes and waiting times) used to interpret the measurement time series, as well as the deconvolution algorithm used to recover the pulse arrival times and amplitudes. In addition, we discuss the quality of the parameter estimation. The results of the density and plasma current scans are presented in section \ref{sec:results}, revealing the fluctuation statistics and demonstrating how the stochastic model parameters change with the Alcator C-Mod plasma parameters. Finally, we discuss the results and conclude the study in section \ref{sec:discussion-conclusions}. \ref{appendix:ip-scan} presents the fluctuation statistics estimated from the plasma current scan in terms of $\Ip$ and \ref{appendix:fit-issues} discusses methodology for estimating model parameters.

\section{Experimental setup}\label{sec:experimental-setup}

\begin{figure}[h!]
    \centering
    \includegraphics[width=0.4\textwidth]{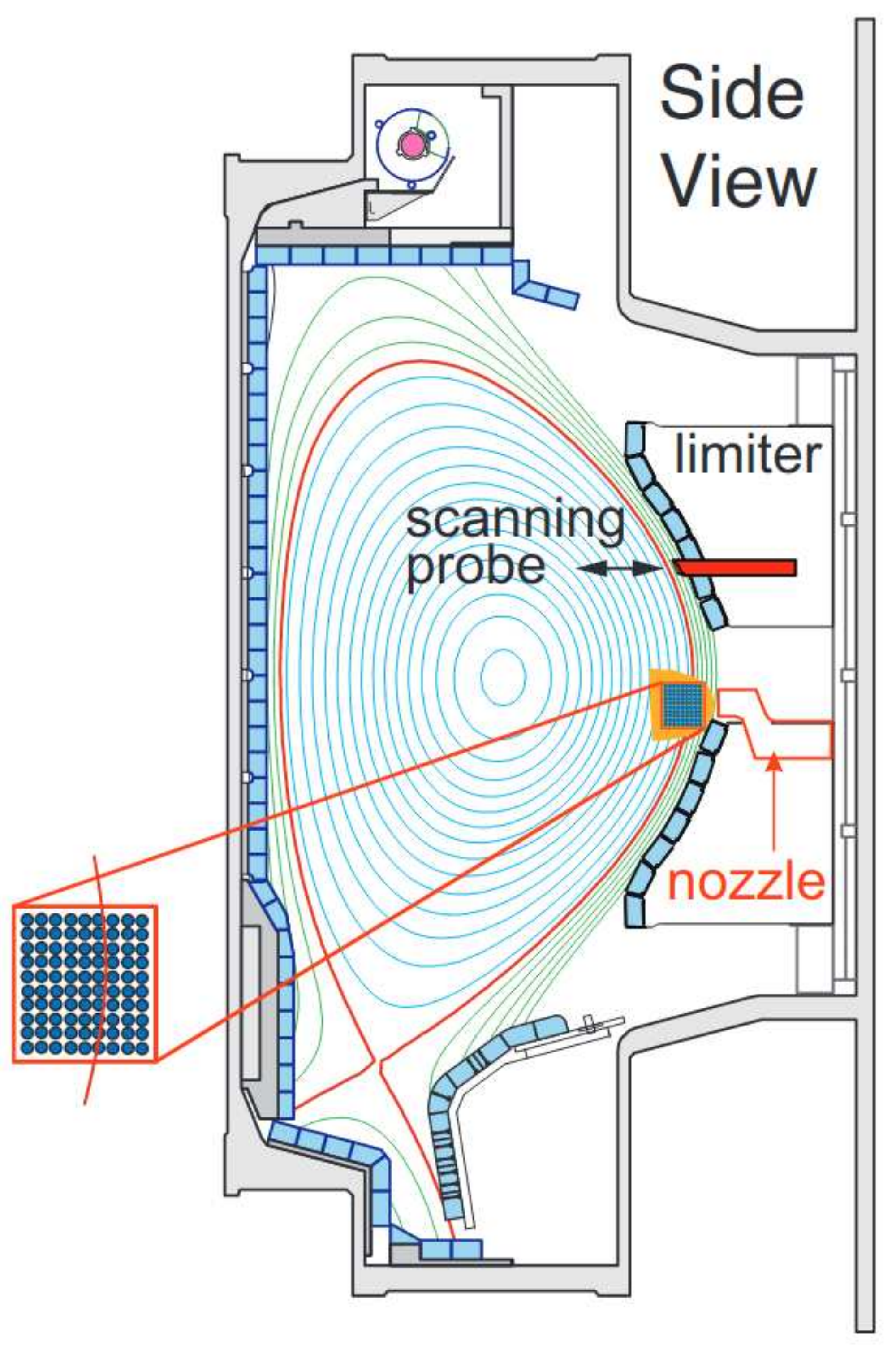}
    \caption{Poloidal cross-section of the Alcator C-Mod tokamak in lower diverted single-null configuration, showcasing the diagnostic set up. The mirror-Langmuir probe (MLP) and the gas puff imaging (GPI) were used to diagnose SOL plasma fluctuations \cite{garcia-2013}. The blow-up of the GPI field-of-view illustrates the $9\times10$ viewing-spots and the LCFS passing through the field-of-view. This figure is reproduced from Garcia O E \textit{et al} in \cite{garcia-2013}, with the permission of AIP Publishing.}
    \label{fig:cmod-cross-section}
\end{figure}

\noindent Alcator C-Mod is a compact high-field (toroidal field $\Bt$ from $2$ to $8\,\text{T}$) tokamak with major radius $R=0.68\,\text{m}$ and minor radius $a=0.21\,\text{m}$. We study the statistical properties of the time series from GPI and the MLP diagnostics measuring far-SOL fluctuations at the outboard side of ohmically heated plasma discharges fuelled by deuterium. All plasma discharges analyzed here were in a diverted LSN magnetic configuration, as shown in figure \ref{fig:cmod-cross-section}.

\subsection{Diagnostic systems}

The GPI diagnostic provides two-dimensional images of emitted line radiation from a neutral gas with high temporal resolution. This diagnostic consists of two essential parts. A gas nozzle puffs a contrast gas into the boundary plasma. The puffed gas atoms are excited by local plasma electrons and emit characteristic line radiation modulated on fast time scales by fluctuations in the local electron density and temperature. This local emission is sampled in a favorable viewing geometry by an optical system comprising either a fast-framing camera or an array of avalanche photodiodes (APDs). 

The APD-based GPI diagnostic on Alcator C-Mod was used for this study, and it consists of a $9 \times 10$ APD array of toroidally-directed views of a localized gas puff. The GPI system was puffing helium and imaging the He I 587 nm emission line, where the signals are digitized at a rate of $2 \times 10^6$ frames per second. The viewing area spans the outboard edge and SOL plasma near the midplane, covering the major radius from $88.00$ to $91.08\,\text{cm}$ and the vertical distance, relative to the $Z=0$ midplane, from $-4.51$ to $-1.08\,\text{cm}$. The in-focus spot size of the imaging optics is $3.8\,\text{mm}$ for each of the individual views. For each discharge analyzed here, the GPI diagnostic yields at least $100$ milliseconds of usable time series data during which both plasma current and line-averaged density were approximately constant. More information on this APD-based GPI system on the Alcator C-Mod device can be found in reference \cite{cziegler-2010}. Since the absolute value of the GPI light intensity is a function of the (unmeasured) local density of neutral gas, it is of secondary significance, and we only consider normalized GPI signals in this study.

The MLP digitization system allows for fast sampling of ion saturation current $\isat$, floating potential $\Vf$, and electron temperature $\Te$ with $0.9\,\mu\text{s}$ time resolution on a single electrode. The plasma parameters are obtained by fitting digitized $I-V$ data, sampled at a rate of $3.3\,\text{MHz}$. Due to its capability of real-time plasma temperature determination to optimize voltage bias states at sampling frequencies in the range of megahertz, it is therefore called a ``mirror" Langmuir probe. Four probe tips are embedded in a Mach probe head that is mounted on a linear servomotor probe drive system. These four MLP electrodes are arranged in a pyramidal dome geometry on the probe head so that they sample approximately the same magnetic flux surface. A diagram of the MLP can be seen in reference \cite{kube-2016-ppcf}, and more details can be found in reference \cite{labombard-2007}. The probe was scanned horizontally or dwelled in a position just inside of the limiter radius $11\,\text{cm}$ above the outboard midplane location. The MLP and the GPI diagnostics did not share the same magnetic flux tube. We focus on the dwell probe measurements in the far-SOL, as we want to investigate the statistical properties of turbulence there. Accordingly, only GPI and scanning probe data from the far-SOL are analyzed in order to make a direct comparison.

\subsection{Data preprocessing}

Measurement data is rarely sampled under perfectly stationary conditions. This may be due to small changes in the LCFS location or small drifts in the main plasma density. Before analyzing the time series, measurements were therefore rescaled to have a locally vanishing mean and unit standard deviation,
\begin{equation}\label{eq:rescaled-signal}
\Snnorm_\mathrm{mv} = \frac{\Sn - \avg{\Sn}_\mathrm{mv}}{\Sn_\mathrm{rms, mv}},
\end{equation}
where
\begin{equation}\label{eq:moving-average}
\avg{\Sn}_\mathrm{mv}(t_i) = \frac{1}{2r+1}\sum_{k=-r}^{r}\Sn(t_{i+k})
\end{equation}
and
\begin{equation}\label{eq:moving-rms}
\Sn_\mathrm{rms, mv}(t_i) = \left[ \frac{1}{2r+1}\sum_{k=-r}^{r} (\Sn(t_{i+k}) - \avg{\Sn(t_i)}_\mathrm{mv})^2\right]^{1/2}
\end{equation}
denote the moving average and the moving root mean square, respectively, of the time series signal $\Sn(t)$ \cite{kube-2018}. For the GPI diagnostic, the sampling time was $\triangle_\text{t} = 0.5\,\mu\text{s}$ and the moving filter radius used was $r = 8192$ samples, which corresponds to $4\,\text{ms}$. For the MLP diagnostic, the sampling time of the fitted $\isat$ data was $\triangle_\text{t} = 0.3\,\mu\text{s}$ where the filter radius was $r = 16384$ samples, which corresponds to a moving window of $5\,\text{ms}$. Detrending the raw time-series measurements according to equation \eqref{eq:rescaled-signal} ensures that slow variations in the mean and variance due to slowly changing plasma conditions are removed. Absorbing these variations into the normalization of the time series allows for the comparison of as many samples as possible to ensure well-converged statistical estimates.

The estimators of the radial electric drift velocity $U$, particle flux $\Gamma_{n}$ and heat flux $\Gamma_{T}$ estimated using the MLP data are given by,
\begin{equation}\label{eq:radial-velocity}
U = \frac{1}{2} \frac{V^{\mathrm{S}}-V^{\mathrm{N}}}{B_\mathrm{MLP}\triangle_{Z}},
\end{equation}
\begin{equation}\label{eq:particle-flux-ne}
\Gamma_{{n}, \wt{n}_\text{e}} = \wt{n}_\text{e}\wt{U},
\end{equation}
\begin{equation}\label{eq:particle-flux-Is}
\Gamma_{{n}, \wt{I}_\text{sat}} = \wt{I}_\text{sat}\wt{U},
\end{equation}
\begin{equation}\label{eq:heat-flux}
\Gamma_{T} = \frac{\Ttm{\Te}}{T_\text{e,rms}}\wt{U}\wt{n}_\text{e} + \frac{\Ttm{\nee}}{n_\text{e,rms}}\wt{U}\wt{T}_\text{e} + \wt{n}_\text{e} \wt{T}_\text{e} \wt{U},
\end{equation}
respectively. Here, $B_\mathrm{MLP}$ is the magnetic field at the position of the probe head and $(V^{\mathrm{S}}-V^{\mathrm{N}})/{\triangle_{Z}}$ is used to estimate the poloidal electric field, where $\Delta_{Z}=2.24\,\text{mm}$ is the separation between the electrodes. `N' and `S' denote the ``north" and ``south" vertically-spaced electrodes, and $V^{\mathrm{N/S}} = ({V_\mathrm{p}}^{\mathrm{NE/SE}}+{V_\mathrm{p}}^{\mathrm{NW/SW}})/2$ \cite{kube-2018}. The particle flux is calculated using detrended and normalized time series according to \eqref{eq:rescaled-signal}, thus postulating that there is no stationary convection in the SOL. $\Gamma_{{n},\wt{I}_\text{sat}}$ and $\Gamma_{{n},\wt{n}_\text{e}}$ denote the particle flux calculated using $\wt{n}_\text{e}$ and $\wt{I}_\text{sat}$ measurements, respectively. For the heat flux measurements, the contribution comes from a convective part, a conductive part and of a part due to triple correlations \cite{kube-2019-nme}. It is noted that the study in reference \cite{kube-2019-nme} investigates the MLP measurements in scanning mode and elucidates the relative contribution of each part to the total heat flux with Greenwald fraction up to $\fg \approx 0.5$ from the same runday on the Alcator C-Mod device. 

\section{Statistical framework}\label{sec:stat-framework}
In this section, we discuss the stochastic model used as the data analysis framework, as well as data processing methods. Implementations of analytical functions and analysis routines can be found in the UiT Complex Systems Modelling group GitHub repository \cite{csm-github}.

\subsection{The stochastic model}

Previous work has shown that the statistical properties of SOL plasma fluctuations in various fusion devices and confinement modes have been accurately described by means of stochastic modeling \cite{theodorsen-2016-ppcf,garcia-2017-nme,garcia-2018}. This stochastic model known as a filtered Poisson process (FPP) describes single-point measurements in the SOL as a super-position of uncorrelated pulses with a fixed shape and duration,
\begin{equation} \label{eq:fpp-def}
    \Sn_K(t) = \sum\limits_{k=1}^{K(T)} A_k \snw\left(\frac{t-s_k}{\taud} \right).
\end{equation}
Here, $K(T)$ is a Poisson process in the interval $[0,T]$ with intensity $T/\Ttm{w}$, where $T$ is the duration of the process and $K$ is the number of pulses. Consequently, pulse arrival times $s_k$ are independent and uniformly distributed on the interval, and waiting times $w_k$ are independent and exponentially distributed with mean value $\Ttm{w}$. All pulses are assumed to have the same duration time $\taud$. The amplitudes $A_k$ are taken to be exponentially distributed with the mean value $\Ttm{A}$. We assume that the pulse function is given by a two-sided exponential,
\begin{equation}  \label{eq:def-exp-two}
  \snw\left(x\right)=
    \begin{cases}  
        \exp(-x/(1-\lambda)), & x\geq0, \\
        \exp(-x/\lambda), & x < 0,
    \end{cases}     
\end{equation}
where the pulse asymmetry parameter is described by $\lambda$ and $x$ is a dimensionless variable. 

The fundamental parameter of the stochastic model is the \textit{intermittency parameter} defined by $\gamma = \taud / \Ttm{w}$, which determines the degree of pulse overlap. When $\gamma$ is small, the pulses appear isolated in the realizations of the process, resulting in a strong intermittency. When $\gamma$ is large, there is significant overlap of pulses, resulting in a weakly intermittent process. For $\gamma \to \infty$, the FPP approaches a normally distributed process \cite{garcia-2012}.  By averaging over all random variables, it can be shown that the four lowest order moments are the mean $\Ttm{\Sn} = \gamma \Ttm{A}$, variance $\rms{\Sn}^2 = \gamma \Ttm{A}^2$, skewness $S_{\Sn} = 2/ \sqrt{\gamma}$ and flatness $F_{\Sn} = 3+6/\gamma$ \cite{garcia-2012}. The PDF, and therefore the moments, do not depend on the pulse asymmetry parameter $\lambda$ \cite{garcia-2016}. It follows that the PDF of $\Sn$ is a Gamma distribution, 
\begin{equation}\label{eq:pdf-signal}
P_{\Sn}(\Sn) = \frac{1}{\avg{A}\Gamma(\gamma)}\left(\frac{\Sn}{\avg{A}}\right)^{\gamma-1}\exp{\left(\frac{\Sn}{\avg{A}}\right)},
\end{equation}
where the shape parameter for the PDF is $\gamma$, the intermittency parameter, $\Gamma$ here denotes the Gamma function and $\avg{A}$ is the scale parameter of the PDF.

Since all experimental measurement signals are normalized to have zero mean and unit standard deviation, the stationary process in \eqref{eq:fpp-def} is scaled to $\Snnorm = (\Sn - \avg{\Sn})/ \Sn_\mathrm{rms}$. The frequency power spectral density (PSD) of the FPP is the product of two Lorenztian spectra, where the analytical expression is \cite{garcia-2017-ac}
\begin{equation}\label{eq:psd-signal}
    \Omega_{\Snnorm}(\omega) = \frac{2\taud}{\left[1+(1-\lambda)^2(\taud \omega)^2\right]\left[1+\lambda^2 (\taud \omega)^2\right]} ,
\end{equation}
where $\omega$ is the angular frequency. The PSD of the normalized process $\widetilde{\Phi}$ is the same as that of a single pulse due to the assumption of independently and uniformly distributed pulse arrivals and fixed pulse duration. The PSD features a flat part for low frequencies and a power-law decay for high frequencies. The intermittency parameter does not influence the shape of the power spectral density \cite{garcia-2017-ac}.

\subsection{Noise and parameter estimation}

Blob dispersion, small background fluctuations and measurement noise may all contribute to deviations from predictions by the FPP for experimental signal values close to the mean value. We model all these fluctuations as an additional normally distributed noise process $X$ with zero mean and variance $\rms{X}^2$. 
We define the noise parameter $\eps$ as the ratio between the root mean square of the noise and the root mean square of the signal, $\eps = \left(\rms{X}/\rms{\Sn}\right)^{2}$. The combined process $\Phi + X$ has mean value $\Ttm{\Sn + X} = \gamma \Ttm{A}$, variance $\rms{(\Sn + X)}^2 = (1+\eps)\gamma \Ttm{A}^2$, skewness $S_{\Sn + X} = 2/\gamma^{1/2}(1+\eps)^{3/2}$ and flatness $F_{\Sn + X} = 3+6/\gamma(1+\eps)^{2}$. 

The PDF of the combined process is a convolution between a Gamma distribution and a normal distribution. By assuming correlated noise which is noise connected to the pulse, the PSD is exactly the same as the expression in \eqref{eq:psd-signal}. If the noise is uncorrelated, also known as observational noise, the PSD is given by equation (35$b$) in reference \cite{theodorsen-2017-ps}. We assume correlated noise and utilize \eqref{eq:psd-signal} for the spectrum in this study. Finally, the expression for the PDF of the normalized signal with the noise parameter can be found in equation (A9) in reference \cite{theodorsen-2017-ps}.

For the MLP measurement data, the shape of the PSD is influenced by the preprocessing, which filters the signal through a 12-point boxcar window \cite{kube-2020}. Figure \ref{fig:density-scan-psd-sig-mlp} shows the resulting PSD. Therefore, the expected PSD of the MLP data time series is the product of the function in equation \eqref{eq:psd-signal} and the PSD of a boxcar window \cite{kube-2020},
\begin{equation}\label{eq:psd-signal-boxcar}
    \Omega_{\Snnorm, \mathrm{MLP}}(\omega) = \Omega_{\Snnorm}(\omega) \left[\frac{1}{6\Delta_\mathrm{t}\omega}\mathrm{sin}(6\Delta_\mathrm{t}\omega)\right]^2.
\end{equation}
The boxcar filtering results in an estimated pulse function that is highly asymmetric with $\lambda$ close to zero.

The parameters of the stochastic model are $\Ttm{A}$, $\taud$, $\lambda$, $\Ttm{w}$ and $\eps$, which can all be estimated from realizations of the process. The parameters $\gamma$ and $\eps$, can be estimated from the PDF of realizations of the process or alternatively from the empirical characteristic function (ECF) of the normalized signal \cite{theodorsen-2017-ps}. The average pulse amplitude $\Ttm{A}$ can then be estimated from $\gamma$ and the sample mean of the realization. The pulse duration $\taud$ and asymmetry parameter $\lambda$ can be estimated from the auto-correlation function or the PSD \cite{garcia-2017-ac}. From $\gamma$ and $\taud$, the average waiting time $\Ttm{w}$ can be estimated. Finally, the deconvolution method can then be used to unravel the pulse arrival times and amplitudes, which allows to estimate the amplitude and waiting time distributions and the mean values of these directly, as will be discussed in section \ref{sec:deconvolution}. Furthermore, in section \ref{subsec:param-analysis} we provide a consistency check between these distribution parameters and the mean values $\Ttm{A}$ and $\Ttm{w}$ estimated from the moments.

\subsection{Deconvolution algorithm}\label{sec:deconvolution}

The RL deconvolution algorithm is a point-wise iterative procedure used to recover the amplitude forcing given a known pulse function \cite{richardson-1972,lucy-1974,benvenuto-2010}. This was done to achieve a broader range of waiting-time and amplitude statistics compared to the conditional averaging method and to relate them to plasma parameters. A detailed description and investigation of the deconvolution method is presented in reference \cite{ahmed-2022}. Figure \ref{fig:flow-chart} shows how the deconvolution algorithm in incorporated into the study.

\tikzstyle{input-output} = [trapezium, trapezium left angle=70, trapezium right angle=110, minimum width = 2cm, minimum height = 0.5cm, text centered, draw=black, fill=blue!20,drop shadow]
\tikzstyle{process} = [rectangle, rectangle, minimum width=3cm, minimum height=0.5cm, text centered, draw=black, fill=yellow!30,drop shadow]
\tikzstyle{arrow} = [thick, ->, >=stealth]

\begin{figure*}[b!]
\centering
\begin{tikzpicture}{node distance=1.5cm}

\node (signal) [yshift=-1cm, label=below:{$\Sn$}] {\frame{\includegraphics[width= 65px]{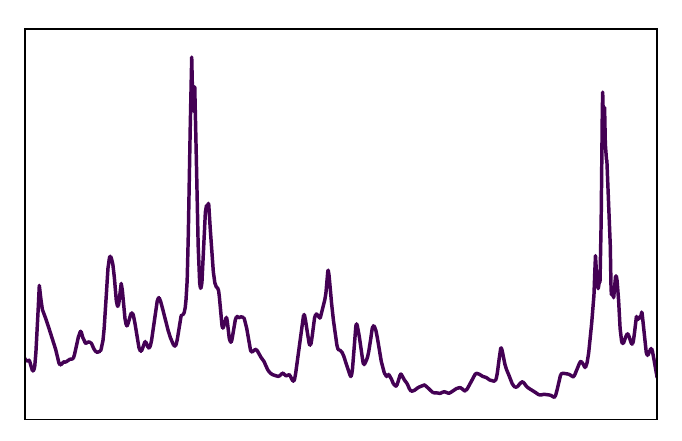}}};
\node (deconvolve_pulse) [right of=signal, xshift=2.5cm, label=below:{$\varphi$}]{\frame{\includegraphics[width= 65px]{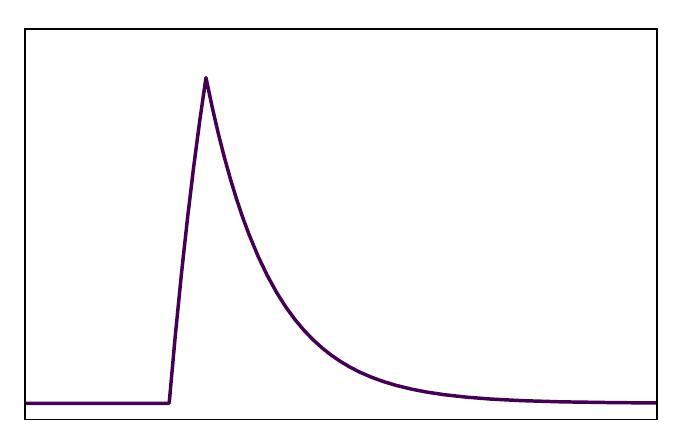}}};
\node (deconvolution) [process, below of=signal, yshift=-1cm, xshift=2cm] {Deconvolution};
\node (forcing_result) [input-output, below of=deconvolution, yshift=-0.5cm, label={[xshift=1.1cm, font=\normalsize\sffamily,name=label1] right:{(a)}}] {$\mathcal{F}$};
\node (peak_finding) [process, right of=forcing_result, xshift = 4.25cm, yshift=1.5cm] {Three-point maxima};
\node (estimated_parameters) [input-output, below of=peak_finding, yshift=-0.75cm] {$A_\mathrm{est}$, $s_\mathrm{est}$};
\node (estimated_w) [input-output, right of=estimated_parameters, xshift=2cm, label={[xshift=0.4cm,font=\normalsize\sffamily,name=label2] right:{(b)}}] {$w_\mathrm{est}$};

\draw [arrow] (signal)--(deconvolution);
\draw [arrow] (deconvolve_pulse)--(deconvolution);
\draw [arrow] (deconvolution)--(forcing_result);
\draw [arrow] (forcing_result) |-([shift={(2.5cm,-3mm)}]forcing_result.south east)-- ([shift={(-0.5cm,3mm)}]peak_finding.north west)-| (peak_finding);
\draw [arrow] (peak_finding)--(estimated_parameters);
\draw [arrow] (estimated_parameters)--(estimated_w);

\begin{scope}[on background layer]
\node[minimum width=6cm,draw,dashed,gray,rounded corners,fill=green!30,fit=(signal) (deconvolve_pulse) (forcing_result)]{};
\node[xshift=0.5cm,minimum width=7cm,draw,dashed,gray,rounded corners,fill=green!30,fit=(peak_finding) (estimated_parameters) (estimated_w)]{};

\end{scope}
\end{tikzpicture}
\caption{A schematic diagram showing how the RL deconvolution is utilized \cite{ahmed-2022}. Firstly, (a) shows a measurement time-series $\Sn$. Here, this is an excerpt of $\Isat$ from a Greenwald fraction of $\fg = 0.80$. The pulse function $\snw$ estimated from the PSD is deconvolved out to estimate $\mathcal{F}$. Finally in (b), a three-point maxima peak-finding algorithm is employed to find the estimated amplitudes $A_\mathrm{est}$, arrival times $s_\mathrm{est}$ and hence the estimated waiting times $w_\mathrm{est}$. The estimated amplitudes and waiting times are then used to estimate their respective distributions.
\label{fig:flow-chart}}
\end{figure*}
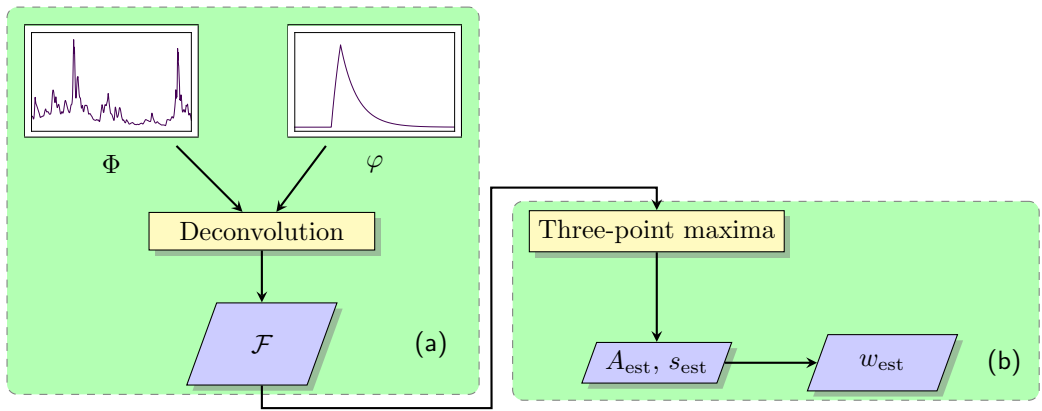

The full (un-normalized) FPP can be expressed as a convolution between the pulse function $\snw$ and a forcing signal $\mathcal{F}_K$,
\begin{equation} \label{eq:signal-conv}
    \Phi_K\left(t\right) = \left[\snw * \mathcal{F}_K\right]\left(\frac{t}{\tau_\mathrm{d}}\right),
\end{equation}
where $\mathcal{F}_K$ consists of a train of delta-function pulses,
\begin{equation} \label{eq:def-fk}
    \mathcal{F}_{K}\left(t\right) = \sum\limits_{k=1}^{K\left(T\right)} A_k \delta\left(\frac{t-s_k}{\tau_\mathrm{d}}\right).
\end{equation}
Given an estimate of $\snw$, we estimate $\mathcal{F}_K$ according to the iterative procedure
\begin{equation} \label{eq:deconv-scheme}
    \mathcal{F}^{(n+1)}_{j} = \mathcal{F}^{(n)}_{j}\frac{\left(\Phi * \wh{\varphi}\right)_j + b}{\left(\mathcal{F}^{(n)} * \varphi * \wh{\varphi}\right)_j + b},
\end{equation}
where the hat symbol $\wh{\cdot}$ is used to denote a flipped vector, $\wh{\varphi}_j = \wh{\varphi}_{-j}$. The parameter $b$ ensures positivity of the deconvolved signal: For $b=0$, positive definite $\Phi$, and positive definite initial guess $\mathcal{F}^{(0)}$, all subsequent iterations $\mathcal{F}^{(n)}$ are also positive definite. To maintain this property in the presence of noise, $b$ is chosen such that $\left(\Sn * \wh{\snw}\right)_j + b > 0\,\forall\,j$ \cite{benvenuto-2010}. The iteration in \eqref{eq:deconv-scheme} is known to converge, minimizing $\Phi - \snw * \mathcal{F}^{(n)}$ in the least-square sense under Gaussian noise. The choice of the initial guess $\mathcal{F}^{(0)}$ as well as the exact value of $b$ may affect the rate of convergence, but does not influence the result as long as $b$ is small compared to the mean signal value. 

The outcome of the deconvolution process yields a time series consisting of pulses that are highly localized. However, it is important to note that the deconvolution does not typically reduce each pulse to the width of a single data point. To recover the pulses and to remove spurious events in parts of the signal without pulses, we apply a simple three-point running maxima with a threshold, tagging each data point as a pulse if it is larger than each of its neighbors and larger than $10^{-3}\Ttm{A}$.\footnote{In reference \cite{ahmed-2022}, a specific threshold relating to $\gamma$ and $\eps$ was used on synthetic realizations of the process with noise which led to exponentially distributed amplitudes and waiting times. We move away from this threshold as it was found to be quite harsh when applied to these experimental measurement time series.} 

The primary focus is to investigate the statistical properties of the fluctuations where the analysis is performed on normalized signals. However, it is not feasible to utilize a normalized signal as input for the deconvolution algorithm. This is because the normalized signal can result in a decay to an incorrect zero level, leading to a distorted representation of the pulse function. To address this issue, we rescale the normalized time series and perform the RL deconvolution algorithm on $\sqrt{\gamma(1+\eps)}\Snnorm + \gamma$. Here, $\gamma$ and $\eps$ are estimated from either the PDF or the ECF of the signal. By incorporating this rescaling approach, we ensure that the deconvolution accurately captures the pulse function without being influenced by the normalization process.

The experimental measurement data reported in the following reveal a bi-exponential distribution of pulse amplitudes and waiting times. Such a bi-exponential pulse amplitude distribution follows from the assumption of a discrete uniform distribution of pulse velocities \cite{losada-2022}. The bi-exponential amplitude distribution is mathematically described as
\begin{equation}\label{eq:bi-exp-amp}
    P_{A}(A) = \frac{q}{\avg{A_\mathrm{<}}}\exp\left(-\frac{A}{\avg{A_\mathrm{<}}}\right) + \frac{1-q}{\avg{A_\mathrm{>}}}\exp\left(-\frac{A}{\avg{A_\mathrm{>}}}\right).
\end{equation}
Here, $0<q<1$ represents the probability that an event corresponds to a small-amplitude fluctuation. $\avg{A_\mathrm{<}}$ denotes the mean of small-amplitude fluctuations, while $\avg{A_\mathrm{>}}$ represents the mean of large-amplitude fluctuations. The average amplitude is given by $\langle{A}\rangle =  q\langle{A_\mathrm{<}}\rangle+(1-q)\langle{A_\mathrm{>}}\rangle$. It is assumed that the amplitudes of these fluctuations correlate with the velocity, where larger-amplitude fluctuations are impacted the least by parallel drainage to the sheaths compared to smaller-amplitude fluctuations \cite{losada-2022}. Consequently, the tail of the signal amplitude distribution is predominantly influenced by the contribution of large-amplitude fluctuations. To estimate the mean values $\langle{A_\mathrm{>}}\rangle$ and $\langle{w_\mathrm{>}}\rangle$, we utilize the expression of the bi-exponential distribution rather than employing a tail fit in order to prevent any imposed hard limits.  These mean values are later compared to $\Ttm{A}$ and $\Ttm{w}$ estimated from the statistical properties of the measurements in section \ref{subsec:param-analysis}.

\section{Results}\label{sec:results}

Here, we present results from the analysis performed on GPI and MLP measurement data from a line-averaged density scan and a plasma current scan. Tables~\ref{tab:gpi-density-table} and~\ref{tab:asp-mlp-density-table} provide details of the duration of time windows considered for the analysis, line-averaged density, plasma current, and toroidal magnetic field considering these time windows, as well as shot numbers. The time windows are chosen such that both the plasma parameters and the fluctuation measurement time series are reasonably stationary. 

\begin{figure}[h!]
    \centering
    \includegraphics[width=0.6\textwidth]{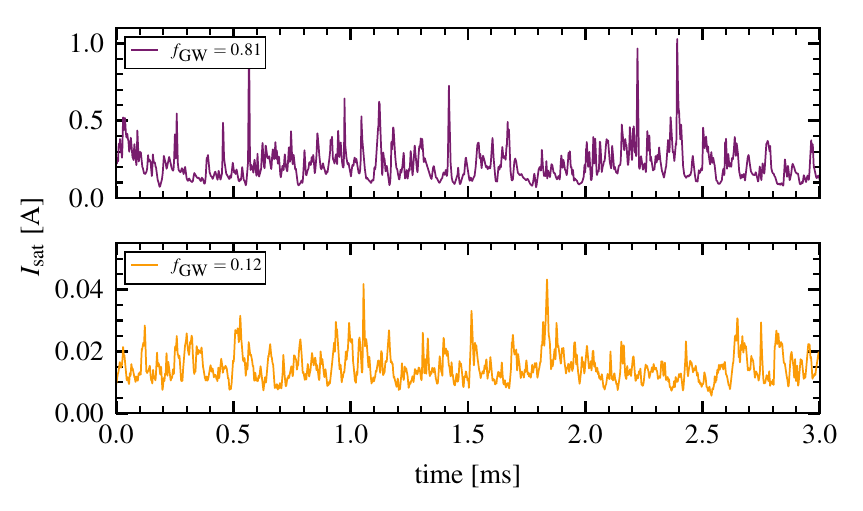}
    \caption{Excerpts of ion saturation current measurements from the MLP dwelling the the far-SOL. The upper panel shows measurements from the highest Greenwald fraction case of the density scan $\fg=0.81$, whereas the lower panel shows the time series from the lowest Greenwald fraction $\fg=0.12$. Note the different scales on the $y$-axes for both time series.
    \label{fig:mlp_time_series}}
\end{figure}

Long time series measurements of at least several hundreds of milliseconds allow us to resolve the tails of the PDF so that $\gamma$ is revealed, as are the flat, low frequency part of the PSDs, which aids in determining $\taud$. In figure \ref{fig:mlp_time_series}, un-normalized excerpts of ion saturation current time series are shown for $\fg = 0.12$ and $\fg = 0.81$. The upper panel of figure \ref{fig:mlp_time_series} shows larger signal amplitudes where the bursts appear more intermittent for $\fg=0.81$ compared to $\fg=0.12$ shown in the lower panel, indicating strongly intermittent, large-amplitude fluctuations at high line-averaged densities. It is worth noting that the background appears to be insignificant compared to the large bursts in both time series.

\begin{table*}[h]
\caption{Plasma discharges considered for the density scan using the GPI fluctuation measurements. We present the start time of the GPI time series analysis $t_\mathrm{start}$ and the duration of the time window considered $T$. In addition, we show the line-averaged densities $\nebar$ and the associated Greenwald fraction $\fg$ averaged over the analysis time window. The plasma current $\Ip$ is in the range $0.51-0.55\,\text{MA}$ and the toroidal magnetic field $\Bt$ is approximately $5.4\,\text{T}$. The time averaged $\rho$-position is in the range $2.3-2.7\,\text{cm}$, where $\rho$ is the radial distance outside the LCFS after magnetically mapping the measurement location to the outboard midplane using EFIT. All of these discharges are from the same runday. These are all LSN except for the last $20-90\,\text{ms}$ in the time series from shots 1160616025 and 1160616026, which are double null according to EFIT. Excluding these parts of the time series did not make a difference to the fluctuation statistics.}
\begin{indented}
\lineup
\item[]\begin{tabular}{@{}*{7}{l}}
\br
shot & $t_\mathrm{start}$ [s] & $T$ [s] & $\nebar$ $[\times10^{20} \, \mathrm{m}^{-3}]$  &$\fg$ \cr
\mr
 1160616009 & 1.25 & 0.2 & 0.86  & 0.24 \\
 1160616011 & 1.15 & 0.3 & 1.07  & 0.30 \\
 1160616016 & 1.15 & 0.3 & 1.60  & 0.45 \\
 1160616017 & 1.15 & 0.3 & 1.56  & 0.45 \\
 1160616018 & 1.15 & 0.3 & 1.65  & 0.47  \\
 1160616022 & 1.15 & 0.3 & 2.29  & 0.67 \\
 1160616025 & 1.15 & 0.3 & 2.76  & 0.82 \\
 1160616026 & 1.15 & 0.3 & 2.83  & 0.85 \\
\br 
 \end{tabular}
 \end{indented}
\label{tab:gpi-density-table}
\end{table*}

\begin{table*}[h!]
\caption{Plasma discharges considered for the density scan using the MLP in dwell mode. We present the start time of the MLP time series analysis $t_\mathrm{start}$ and the duration of the time window considered $T$. In addition, we show the same parameters that are given in table \ref{tab:gpi-density-table}, together with the time-averaged position of the probe dwelling to take these measurements, which are quoted in terms of the relative distance from the LCFS, $\rho$. In machine coordinates, the probe is dwelling in the range $86.4-86.9\,\text{cm}$ across all of these plasma discharges. The plasma current $\Ip$, is in the range $0.51-0.56\,\text{MA}$ and the toroidal magnetic field $\Bt$ is approximately $5.4\,\text{T}$. All of these discharges are from the same runday as in table \ref{tab:gpi-density-table}. The connection length for the density scan is in the range $8-10\,\text{m}$. All of these plasma discharges were in LSN.}
\begin{indented}
\lineup
\item[]\begin{tabular}{@{}*{8}{l}}
\br
shot & $t_\mathrm{start}$ [s] & $T$ [s] & $\nebar$ $[\times10^{20} \, \mathrm{m}^{-3}]$ &$\fg$ & $\rho$ [cm] \cr
\mr
 1160616007 & 1.03 & 0.27 & 0.46 & 0.12 & 1.23\\
 1160616008 & 0.83 & 0.67 & 0.45 & 0.13 & 1.05\\
 1160616010 & 0.5 & 1.0 & 0.76  & 0.21 & 1.25\\
 1160616012 & 0.7 & 0.8 & 1.00 & 0.27 & 1.41\\
 1160616015 & 0.75 & 0.35 & 1.29  & 0.36 & 1.40\\
 1160616019 & 0.7 & 0.8 & 1.58  & 0.45 & 1.38 \\
 1160616018 & 0.7 & 0.8 & 1.63  & 0.47 & 1.28 \\
 1160616021 & 0.65 & 0.4 & 1.86  & 0.58 & 1.02 \\
  1160616020 & 1.0 & 0.5 & 2.12  & 0.60 & 1.17 \\
 1160616023 & 1.0 & 0.5 & 2.05  & 0.62 & 0.97 \\
 1160616024 & 1.0 & 0.5 & 2.09  & 0.63 & 0.76 \\
1160616027 & 1.0 & 0.35 & 2.70  & 0.80 & 0.86 \\
 1160616026 & 1.0 & 0.35 & 2.73  & 0.81 & 0.88 \\
 \br 
 \end{tabular}
 \end{indented}
\label{tab:asp-mlp-density-table}
\end{table*}

While the data for the scan in particle density comes from a dedicated experiment executed on a single runday, the data for the current scan were gathered from other experiments on different rundays and under conditions over which the line-averaged density was not exactly the same. Tables~\ref{tab:gpi-current-scan-table} and~\ref{tab:asp-mlp-current-scan-table} show the plasma parameters considered for this scan, for the GPI and MLP data respectively.

For the current scan, the MLP was operated in scanning mode, and a rather large radial bin in the far-SOL was chosen in order to calculate relevant statistical averages and distributions. Therefore, the time windows used on the MLP in this plasma current scan are shorter than the ones used in the other parameter scans, giving larger uncertainty in the parameter estimation and deconvolved amplitudes and waiting times in particular.

\begin{table*}
\caption{Plasma discharges considered for the plasma current scan using the GPI fluctuation measurements. We present the start time of the GPI time series analysis $t_\mathrm{start}$, the length of the time window considered $T$, the plasma current $\Ip$ and toroidal magnetic field $\Bt$. Divertor configurations for these discharges were all in LSN.} 

\begin{indented}
\lineup
\item[]\begin{tabular}{@{}*{7}{l}}
\br     
shot & $t_\mathrm{start}$ [s] & $T$ [ms] & $\nebar$ $[\times10^{20} \mathrm{m}^{-3}]$ & $\Ip$ $[\mathrm{MA}]$& $\Bt$ $[\mathrm{T}]$ &$\fg$ \cr
 \mr
1160629026 & 1.22 & 0.28 & 1.87 & 1.07 & 5.40 & 0.27 \cr
 1160629031 & 1.22 & 0.28 & 1.88 & 1.07 & 5.40 & 0.27 \cr
 1160927003 & 0.78 & 0.61 & 1.44 & 0.79 & 5.36 & 0.28 \cr
 1160616017  & 1.15 & 0.30 & 1.56 & 0.53 & 5.41 & 0.45 \cr
 1160616016 & 1.15 & 0.30 & 1.60 & 0.53 & 5.40 & 0.46 \cr
 \br
 \end{tabular}
\end{indented}
\label{tab:gpi-current-scan-table}
\end{table*}

\begin{table*}
\caption{Plasma discharges considered for the plasma current scan using the ion saturation current data from the scanning MLP. We present the start time of the GPI time series analysis $t_\mathrm{start}$, the length of the time window considered $T$, the plasma current $\Ip$ and toroidal magnetic field $\Bt$. Divertor configurations for these discharges were all in LSN.}

\begin{indented}
\lineup
\item[]\begin{tabular}{@{}*{7}{l}}
\br  
shot & $t_\mathrm{start}$ [s] & $T$ [ms] & $\nebar$ $[\times10^{20} \mathrm{m}^{-3}]$ & $\Ip$ $[\mathrm{MA}]$& $\Bt$ $[\mathrm{T}]$ &$\fg$ \cr
 \mr 
  1160629031 & 1.3011 & 20 & 1.74 & 1.07 & 5.41 & 0.25  \cr
 1140730018 & 1.1869 & 13 & 1.60 & 0.80 & 5.40 & 0.27  \cr
 1160616016 & 1.2583 & 14 & 1.62 & 0.53 & 5.40 & 0.46  \cr
 \br
 \end{tabular}
 \end{indented}
\label{tab:asp-mlp-current-scan-table}
\end{table*}

The GPI signals of interest were taken from diode view positions in the far-SOL. The time windows for the analyses were chosen so that the plasma parameters were steady and to ensure sufficient duration of the time series for estimating the FPP model parameters and distributions. 

To assess the quality of the parameter fitting, we take samples from several GPI view positions at similar radial positions $\rho$ in the far-SOL. Here, $\rho$ is the radial distance outside the LCFS after magnetically mapping the measurement location to the outboard midplane by applying the magnetical equilibrium reconstruction calculated using EFIT. The position of the LCFS is thus at $\rho=0$. The position of the LCFS may change slowly relative to the fixed locations of the views; therefore, the flux surfaces may move slightly relative to the GPI views. However, we will only consider fixed view positions. 

We emphasize that the positions of the GPI and MLP in dwell mode are measuring fluctuations at different $\rho$ values, where the MLP measurements are inside of those of the GPI. Power balance correction was applied to the GPI and MLP $\rho$ positions in all of the results shown in the study. Such corrections were made in order to mitigate possible EFIT errors in the location of the LCFS \cite{labombard-2014}.

We focus mainly on the analysis performed on the ion saturation current from the MLP, as this is a widely measured plasma quantity across various devices. We point the reader to references \cite{kube-2018} and \cite{kube-2019-rsi}, where data analysis has been performed on $\nee$ and $\Te$ fluctuation measurements.

\subsection{Radial profiles}

\begin{figure}[h!]
    \centering
    \subfigure[]{\includegraphics[width=0.48\textwidth]{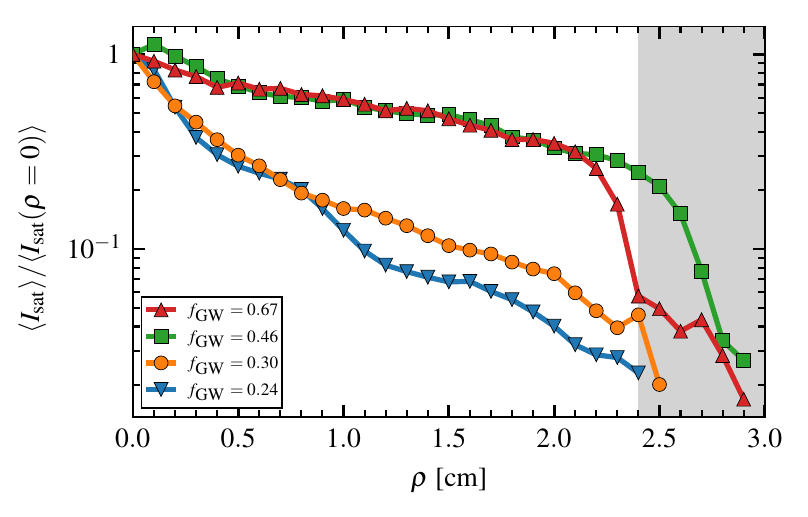}
    \label{fig:asp-mlp-density-profiles-mean-Is}}
    \subfigure[]{\includegraphics[width=0.48\textwidth]{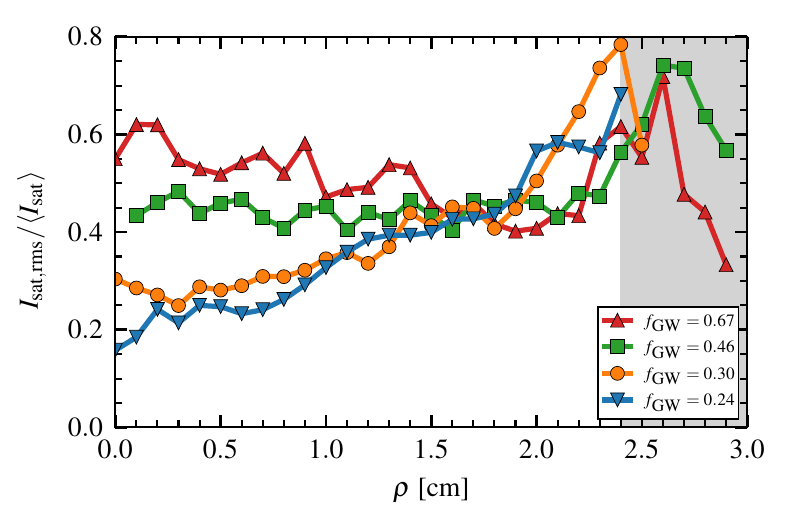}\label{fig:asp-mlp-density-profiles-rlf-Is}}\\
    \caption{Radial profiles of plasma parameters from the density scan: (a) $\avg{\isat}$ normalized by its estimated separatrix value and (b) $\isat$ relative fluctuations. The local limiter location at $Z=11\,\text{cm}$ above the midplane is $R=88.4\,\text{cm}$ (in local machine coordinates). The approximate flux-position location of the limiter using EFIT magnetic reconstruction is presented by the gray-shaded region.}
    \label{fig:asp-mlp-density-profiles}
\end{figure}

\begin{figure}[ht]
    \centering
    \subfigure[]{\includegraphics[width=0.48\textwidth]{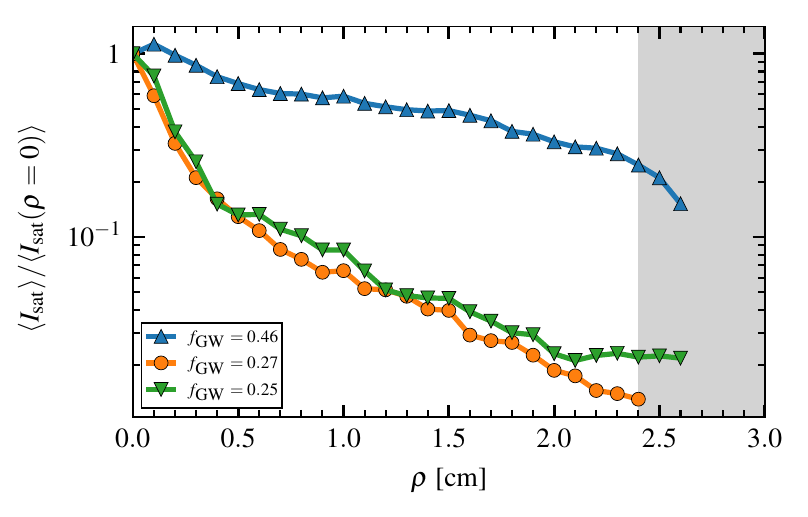}\label{fig:asp-mlp-current-scan-profiles-mean-Is}}
    \subfigure[]{\includegraphics[width=0.48\textwidth]{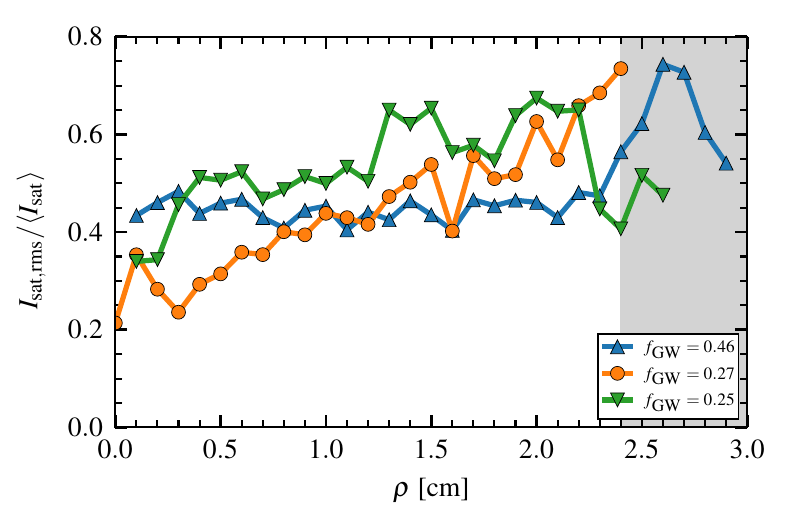}\label{fig:asp-mlp-current-scan-profiles-rfl-Is}}\\
    \caption{Radial profiles of plasma paramters from the plasma current scan: (a) $\avg{\isat}$ normalized by its estimated separatrix value and (b) $\isat$ relative fluctuation levels. The local limiter location at $Z=11\,\text{cm}$ above the midplane is $R=88.4\,\text{cm}$. We show the approximate $\rho$ location of the limiter using EFIT magnetic reconstruction represented by the gray-shaded region.}
    \label{fig:asp-mlp-current-scan-profiles}
\end{figure}
The time-averaged radial profiles of the $\isat$ measurements from the scanning MLP in the density scan are presented in figure \ref{fig:asp-mlp-density-profiles}. These were averaged over $1\,\text{mm}$ radial bins. The profiles are normalized to their respective separatrix values. The location of the limiter shadow, which is always fixed in major radius coordinates, is represented by the gray-shaded region in the $\rho$-coordinate space. When presenting this position relative to the LCFS, the $\rho$ coordinate could change by approximately $5\,\text{mm}$ during the pulse as noted above. For that reason, we show an approximate flux-position location of the limiter in figures \ref{fig:asp-mlp-density-profiles-mean-Is} and \ref{fig:asp-mlp-density-profiles-rlf-Is} as a gray-shaded region that is not fixed relative to the LCFS. In machine coordinates, the probe was always $2\,\text{cm}$ inside of the limiter radius for all discharges in the density scan. The time-averaged radial profiles in figures \ref{fig:asp-mlp-density-profiles-mean-Is} and \ref{fig:asp-mlp-current-scan-profiles-mean-Is} are normalized by the separatrix value to emphasize the profile shape. However, as seen later in figure \ref{fig:asp-mlp-mean}, there is no discontinuous jump in the mean values of the far-SOL quantities, $\isat$, $\nee$ and $\Te$, over the full density and $\fg$ scan.

Focusing on the lower Greenwald fraction cases in figure \ref{fig:asp-mlp-density-profiles-mean-Is}, a well-known two-layer structure can be seen from the radial profile of $\avg{\isat}$. Closer to the separatrix, the near-SOL region shows a steep decay length with moderate fluctuation levels for these measurements. In comparison, the far-SOL exhibits longer scale lengths and a fluctuation level of order unity. As the line-averaged density increases, for Greenwald fractions $\fg \geq 0.46$, the far-SOL profile becomes broader and flatter, so that the far-SOL profile effectively extends all the way to the separatrix. The significant change in the mean profile can be attributed to the amplitudes becoming larger as well as intermittent, which impacts the nature of the cross-field transport in the SOL.

The radial variation in the $\isat$ relative fluctuations levels is presented in figure \ref{fig:asp-mlp-density-profiles-rlf-Is}. Once again, the time-averaged quantities were calculated over $1\,\text{mm}$ bins. The $\isat$ relative fluctuation levels are estimated as the ratio between the standard deviation and the sample mean. Near the separatrix, the relative fluctuation levels are low, resulting in significant pulse overlap for $\fg \leq 0.46$. For $\fg$ = 0.24 the $\isat$ relative fluctuation levels increase from $\sim$ 0.15 to 0.6 from the LCFS into the far-SOL, but are considerably higher for $\fg$ = 0.67 and are around 0.5 over the entire SOL. The far-SOL scale lengths for radial $\avg{\isat}$  profiles were found to be $2.8\,\text{cm}$ for $\fg=0.67$, $2.2\,\text{cm}$ for $\fg=0.46$, $1.2\,\text{cm}$ for $\fg=0.30$ and $1.0\,\text{cm}$ for $\fg=0.24$, hence these scale lengths getting longer with line-averaged density. Overall, this suggests that for the highest line-averaged density studied, the whole SOL is dominated by large-amplitude fluctuations, suggesting that the cross-field transport comes mainly from the filaments.

These time-averaged radial profiles and relative fluctuation levels of the electron density and electron temperature have been previously reported in reference \cite{kube-2019-nme} for $0.1\leq\fg\leq0.5$, and are therefore not shown here. The mean $\nee$ radial profile behaves similarly to the $\isat$ mean profile in figure \ref{fig:asp-mlp-density-profiles-mean-Is}. The mean $\Te$ profiles decay strongly for the highest-density case, indicating that at higher densities the temperature drains faster. For low densities, this is similar to the time-averaged $\isat$ radial profile. The $\nee$ relative fluctuation levels for the same discharges in figure \ref{fig:asp-mlp-density-profiles} were found to vary little with radial distance for all Greenwald fractions. Furthermore, the $\Te$ relative fluctuation levels were found to be consistently higher for $\fg \geq 0.46$ across the entire SOL compared to $\fg \leq 0.30$ but once again, varies little with radial position. Radial profiles of the relative fluctuation levels for the Greenwald fraction ranges $0.15\leq\fg\leq0.30 $ have been previously demonstrated to increase radially outward for all line-averaged densities using the GPI \cite{theodorsen-2017-nf}.

The time-averaged radial profiles for the $\isat$ measurements are presented in figure \ref{fig:asp-mlp-current-scan-profiles} for various plasma currents and these are quoted in terms of Greenwald fractions. At large $\Ip$ values ($\fg=0.25$ and $\fg=0.27$), the distinct two-layer structure between the near- and far-SOL is again obvious, as was observed in figure \ref{fig:asp-mlp-current-scan-profiles-mean-Is}, but there is no significant difference between these two profiles. The $\isat$ mean profile shows a broad and flat profile at the lowest $\Ip$ ($\fg=0.46$), decreasing to 70 \% of the reference value $\avg{\isat(\rho = 0)}$ at the limiter, as shown in figure \ref{fig:asp-mlp-current-scan-profiles-mean-Is}. This is the same data from the same probe reciprocation as the one shown in figure \ref{fig:asp-mlp-density-profiles-mean-Is}. The far-SOL scale lengths from the radial $\avg{\isat}$ profiles were found to be $2.4\,\text{cm}$ for $\fg=0.46$, $0.9\,\text{cm}$ for $\fg=0.27$ and $1.0\,\text{cm}$ for $\fg=0.25$, therefore a decrease in the far-SOL scale length as the $\Ip$ is increases. The radial profiles with mean values of $\nee$ and $\Te$ show similar behavior to that for $\isat$ displayed in figure \ref{fig:asp-mlp-current-scan-profiles}. 

We now focus on the relative fluctuation levels of $\isat$ in figure \ref{fig:asp-mlp-current-scan-profiles-rfl-Is}. Here, $\fg =0.27$ (high $\Ip$) is comparable to both $\fg = 0.24$ and $\fg = 0.30$ in figure \ref{fig:asp-mlp-density-profiles-rlf-Is}. Again $\fg = 0.46$ is the same in both figures \ref{fig:asp-mlp-density-profiles-rlf-Is} and \ref{fig:asp-mlp-current-scan-profiles-rfl-Is}. The highest $\Ip$ case, $\fg = 0.25$, behaves differently, but this discrepancy does not seem to affect the far-SOL statistics discussed in section \ref{subsec:param-analysis}. It does, however, prevent us from drawing firm conclusions about the effect of high $\Ip$ on the SOL profiles.


\subsection{Fluctuation statistics}\label{sec:density-fluc-stats}
We present a detailed analysis of PDFs and PSDs for $\Isat$ signals from the MLP when dwelling in the far-SOL and for the GPI signals across the density scan. For this parameter scan we have long time series measurements from both diagnostics. We will focus on three different density/Greenwald-fraction cases, $\fg=0.24$, $0.67$ and $0.85$ for the GPI measurements and $\fg=0.21$, $0.47$ and $0.80$ for the MLP measurements. For simplicity and ease of interpretation, the figures of PDFs and PSDs from GPI will focus on time series measurements from the same APD view $(R,Z) = (90.68, -1.57)\,\text{cm}$ in machine coordinates, where $\rho=2.3\,\text{cm}$ for $\fg=0.24$, $\rho=2.4\,\text{cm}$ for $\fg=0.67$ and $\rho=2.6\,\text{cm}$ for $\fg=0.85$. Figures showing the parameters of the stochastic model as a function of all Greenwald fractions showing all the views considered are presented in section \ref{subsec:param-analysis}. 

\begin{figure*}[h!]
    \centering
    \subfigure[]{\label{fig:density-scan-pdf-sig-gpi}\includegraphics[width=0.48\textwidth]{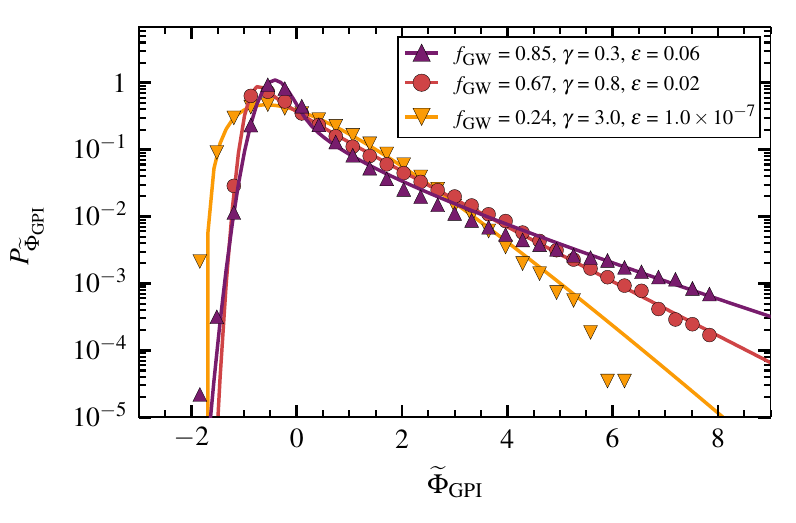}}
    \subfigure[]{\label{fig:density-scan-pdf-sig-mlp}\includegraphics[width=0.48\textwidth]{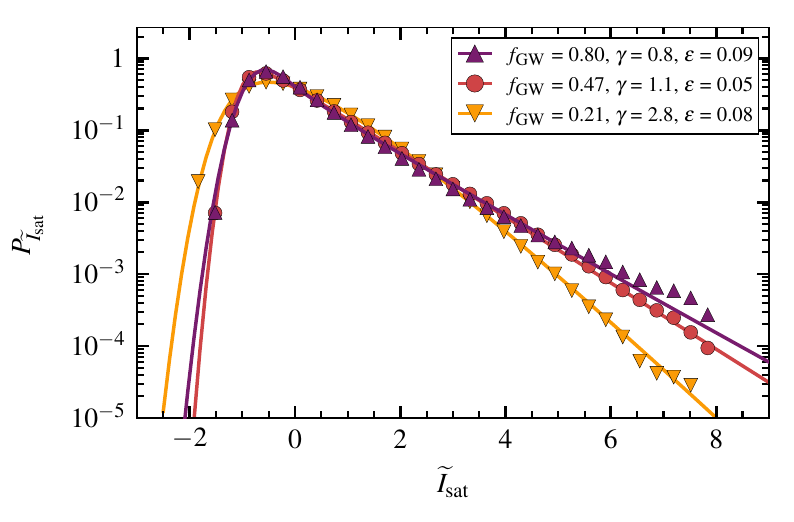}}
    \caption{The PDFs of the normalized time series of (a) the GPI light intensity measurements and (b) the MLP $\Isat$ fluctuation measurements for various line-averaged densities. The markers represent the measurement data. The solid lines represent the corresponding agreement of the measurement data with the stochastic model where the values for the intermittency and the noise-to-signal ratio $\eps$, are quoted.}
    \label{fig:density-scan-pdf-sig}
\end{figure*}

\paragraph{Probability distributions}
We present the histograms of the time-series measurements in figure \ref{fig:density-scan-pdf-sig}. The PDFs of the GPI measurement data for various densities are shown in figure \ref{fig:density-scan-pdf-sig-gpi}, while the MLP measurements are exhibited in figure \ref{fig:density-scan-pdf-sig-mlp}. The parameters $\gamma$ and $\eps$ are estimated from the ECF of the normalized time series \cite{theodorsen-2017-ps}. In all cases, the PDFs are positively skewed and flattened, indicating intermittent fluctuations even at low densities. The tails of the PDFs lift as the density increases for both GPI and MLP measurements, indicating increasingly intermittent time series. The noise ratio $\eps$ is low for all cases considered; the highest attained values are $0.12$ and $0.08$ for the MLP and GPI, respectively. Both of these maximal values were attained at the highest density.

\begin{figure*}
    \centering
    \subfigure[]{\label{fig:density-scan-psd-sig-gpi}\includegraphics[width=0.48\textwidth]{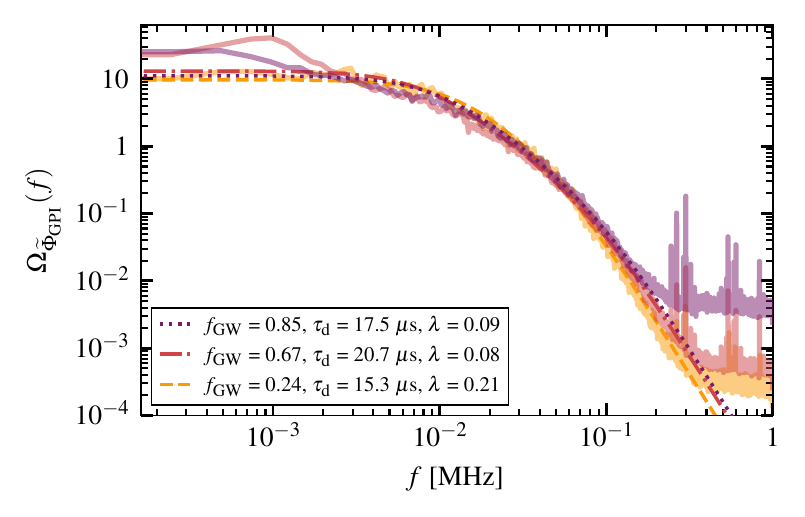}}
    \subfigure[]{\label{fig:density-scan-psd-sig-mlp}\includegraphics[width=0.48\textwidth]{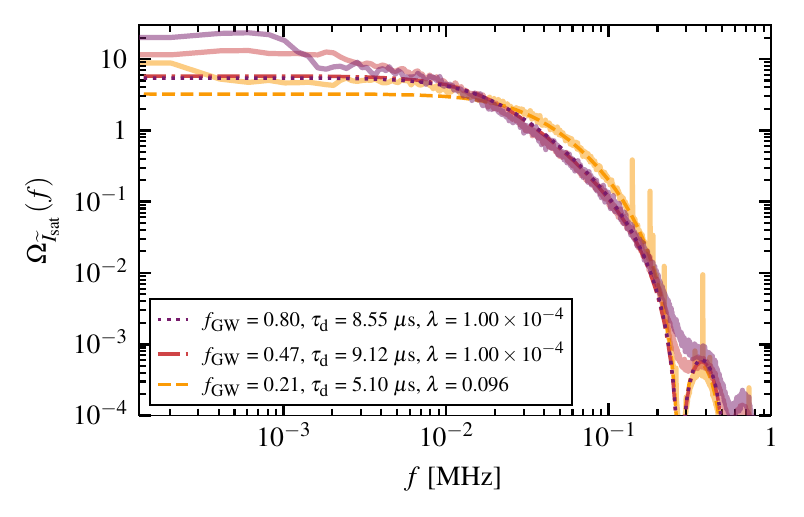}}
    \caption{The PSDs of the normalized time series from (a) the GPI and (b) the MLP ion saturation current measurements for various line-averaged densities. The measurement datasets are represented by the solid lines with lighter colors whereas best fits of the stochastic model are represented by the textured lines of the same colors. \ref{appendix:fit-issues} discusses how these fits were made.}
    \label{fig:density-scan-psd-sig}
\end{figure*}

\paragraph{Power spectral densities}
As noted in section \ref{sec:stat-framework}, the shape of the pulse function is reflected in the frequency dependence of the power spectral density. Assuming a two-sided exponential pulse function as described by \eqref{eq:def-exp-two}, the parameters $\taud$ and $\lambda$ can be estimated from the PSD of the normalized time series. The PSDs of the GPI and MLP time series for various line-averaged densities are presented in figures \ref{fig:density-scan-psd-sig-gpi} and \ref{fig:density-scan-psd-sig-mlp}, respectively. The fits in figure \ref{fig:density-scan-psd-sig-gpi} use \eqref{eq:def-exp-two} directly, while the fits in figure \ref{fig:density-scan-psd-sig-mlp} use the spectrum of the pulse function convolved by the 12-point boxcar, shown in \eqref{eq:psd-signal-boxcar}. As seen previously \cite{garcia-2018, theodorsen-2017-nf}, the power spectra of the GPI time series collapse to a similar shape. At $\fg=0.85$, the relative noise floor of the spectra, evident above about 300 kHz, increases significantly. This apparent increase in the noise floor is actually due to the signal normalization, as seen in \eqref{eq:rescaled-signal}, and the fact that the overall GPI signal is lower at the highest densities relative to the electronic noise. This apparent increase in noise is consistent with the highest density having the highest $\eps$-value, as discussed in the previous paragraph. For the MLP spectra, the ringing effect at high frequencies is clearly seen due to the preprocessing of the MLP data. The lowest density case has a significantly shorter duration than the higher density cases, as is visible in the spectra. 

It is worth noting the differences in the estimates $\taud$ and $\lambda$ of the two diagnostics. The pulse asymmetry parameter estimated from the GPI measurement of the light fluctuations is larger compared to the asymmetry parameter estimated from the MLP $\Isat$ measurements. Extremely small $\lambda$ values estimated for the MLP for $\fg>0.21$ seem to be hitting the lower limit of the fitting function. Overall, $\lambda$ seems to get smaller as the density increases revealing highly asymmetric average pulse shapes. Furthermore, the $\taud$ estimates of the MLP measurements are smaller compared to those of the GPI. This will be discussed further in connection with figure \ref{fig:td-vs-fgw}.
 
\begin{figure}
    \centering
    \subfigure[]{\includegraphics[width=0.48 \textwidth]{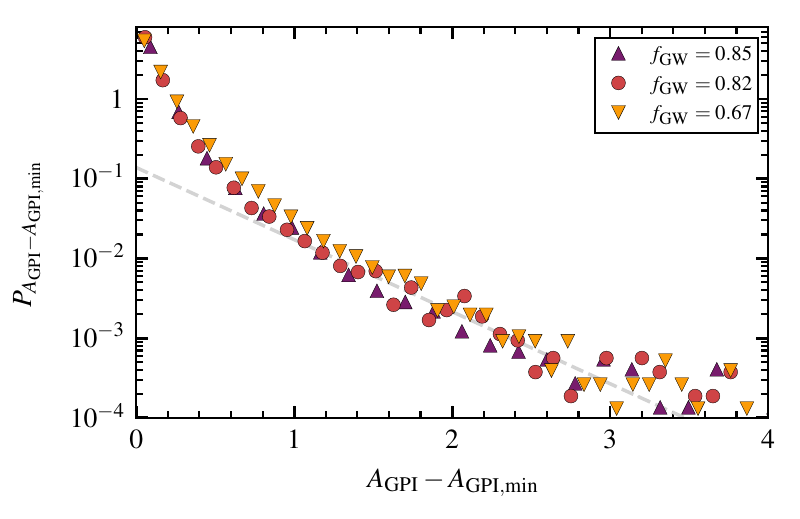} \label{fig:gpi-density-pdf-amp}}
    \subfigure[]{\includegraphics[width=0.48\textwidth]{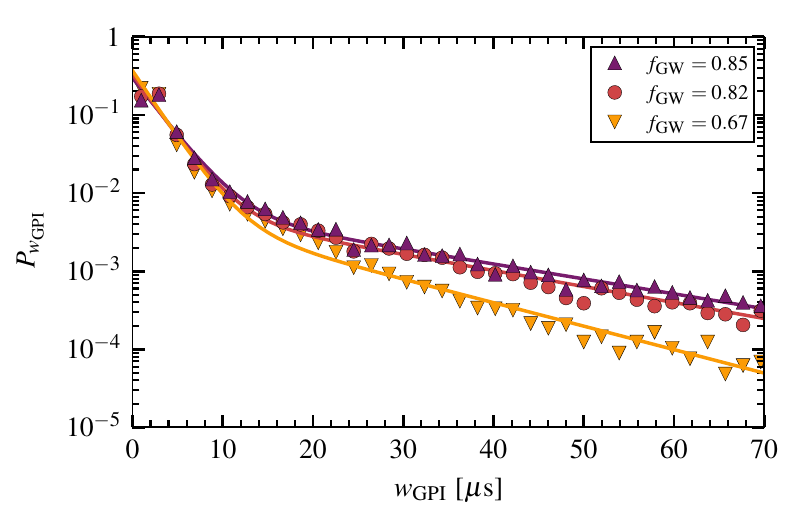}\label{fig:density-scan-pdf-wait-gpi}}
    \caption{Estimated distributions from the deconvolution algorithm applied to the normalized GPI time series. (a) Amplitude distribution of the normalized time series and (b) shows the waiting time distribution. The markers represent the histogram of the measurements for various densities. The gray-dashed line in (a) highlights an exponential decay. The solid lines show the bi-exponential fits to the waiting time distributions.\label{fig:density-scan-pdf-amp-wait-gpi}}
\end{figure}

\begin{figure}
    \centering
    \subfigure[]{\includegraphics[width=0.48\textwidth]{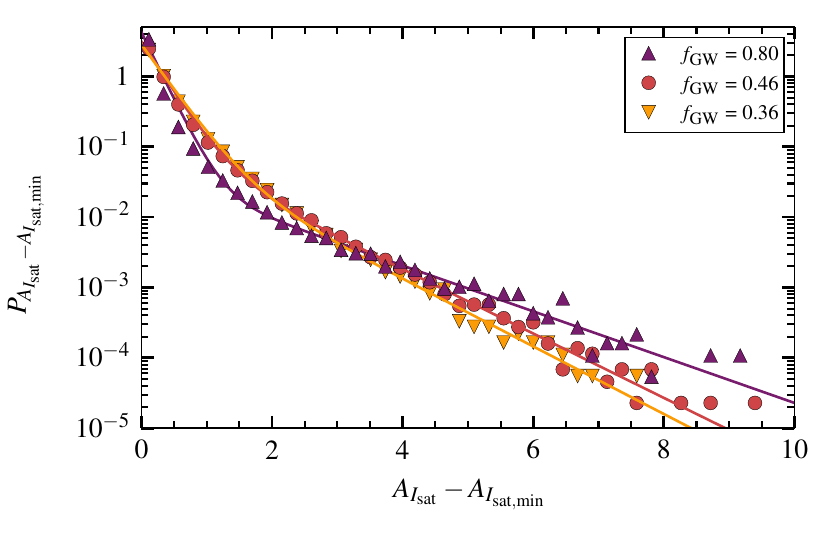}\label{fig:asp-mlp-density-pdf-amp-Is}}
     \subfigure[]{\includegraphics[width=0.48\textwidth]{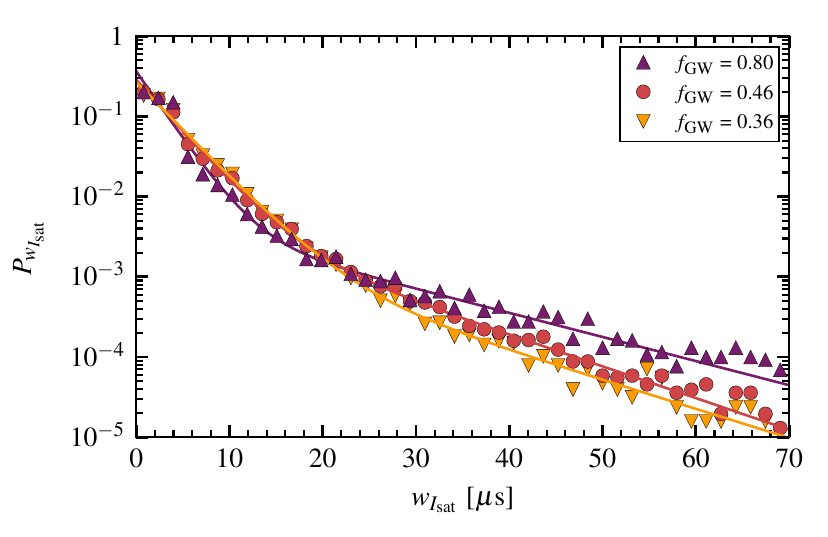}\label{fig:density-scan-pdf-wait-mlp}}
    \caption{Estimated distributions from the RL deconvolution applied to the normalized MLP ion saturation current measurements $\widetilde{I}_\mathrm{sat}$, where (a) shows the amplitudes and (b) are the waiting times for various densities. The markers in the legend represent the measurement data whereas the solid lines show the bi-exponential fits to these waiting time distributions.}
    \label{fig:density-scan-pdf-amp-wait-mlp}
\end{figure}

\paragraph{Deconvolved pulse amplitude distributions}
The pulse amplitude and waiting time distributions estimated from the deconvolution algorithm for various line-averaged densities are presented in figures \ref{fig:density-scan-pdf-amp-wait-gpi} and \ref{fig:density-scan-pdf-amp-wait-mlp}. Here, we show some of the Greenwald density fraction discharges meeting the criterion of deconvolution where $\gamma\theta \leq 1/20$, where $\theta = \triangle_\text{t}/\taud$ is the sampling time normalized by the duration time \cite{ahmed-2022}. These means were estimated by performing a bi-exponential fit to the estimated amplitude distribution using \eqref{eq:bi-exp-amp} and using the exponential that describes the larger events. Further details of this can be found in \ref{appendix:fit-issues} describing the challenges with fitting a bi-exponential to the GPI amplitude distribution. A cross-correlation analysis between pulse amplitudes and both preceding and following waiting times reveals that there are no such correlations, consistent with the assumption of uncorrelated pulses in the FPP model.

We present the estimated distributions from the deconvolution algorithm applied to the normalized GPI time series in figure \ref{fig:density-scan-pdf-amp-wait-gpi}. The estimated amplitude distributions for various Greenwald fractions on the GPI time series are presented in figure \ref{fig:gpi-density-pdf-amp}. The lack of a readily apparent density dependence can be attributed to the fact that the amplitude distribution depicted is derived from normalized signals. The deviation for large amplitudes from an exponential is due to the single data points in the histograms. The relative scarcity of data points did not allow high confidence in the bi-exponential fits to the GPI amplitudes, hence we do not quote their mean values as well as their fits. Instead, we show a gray-dashed line through the large amplitudes to emphasize the exponential decay. In figure \ref{fig:density-scan-pdf-wait-gpi}, the estimated waiting time distribution for the GPI is shown. The estimated mean waiting times from the GPI measurements $\avg{w_\mathrm{>,GPI}}$ are $14.4\,\mu\text{s}$ for $\fg=0.67$, $21.8\,\mu\text{s}$ for $\fg=0.82$ and $22.7\,\mu\text{s}$ for $\fg=0.85$. This shows that as the line-averaged density is increased, the waiting times on average are becoming longer and the fluctuations are becoming increasingly intermittent in the GPI signal.

The RL deconvolution was applied to $\widetilde{I}_\mathrm{sat}$ measurements where the results are shown in figure \ref{fig:density-scan-pdf-amp-wait-mlp}. The functional shape seen in figure \ref{fig:asp-mlp-density-pdf-amp-Is} indicates a density dependence in the amplitude distributions as the density limit is approached. The pulse amplitudes are approximately bi-exponentially distributed for all signals analyzed. For the distributions of the Greenwald fraction cases shown here, the mean of the estimated large amplitudes for the $\wt{I}_\mathrm{sat}$ measurements $\avg{A_{>,\isat}}$ in dimensional units are $0.09\,\text{A}$ for $\fg=0.36$, $0.12\,\text{A}$ for $\fg=0.46$ and $0.29\,\text{A}$ for $\fg=0.80$. In section \ref{subsec:param-analysis}, we show a consistency check of these means to $\Ttm{A}$ estimated from the sample mean $\Phi/\gamma$. For the MLP $\Isat$ measurements, the estimated waiting time distribution is shown in figure \ref{fig:density-scan-pdf-wait-mlp}. The mean of the estimated waiting times $\avg{w_{>,\isat}}$ for these densities are $11.7\,\mu\text{s}$ for $\fg=0.36$, $11.3\,\mu\text{s}$ for $\fg=0.46$ and $15.4\,\mu\text{s}$ for $\fg=0.80$. This shows that as the density limit is approached, longer waiting times are expected due to increasingly intermittent fluctuations.

\subsection{Parametric analysis}\label{subsec:param-analysis}

Here, we investigate how fluctuating quantities measured by the MLP change in the line-averaged density scan in order to explain later the overall results from the estimations of the stochastic model's parameters.

\begin{figure*}[htb!]
    \centering
    \subfigure[]{\includegraphics[width=0.48\textwidth]{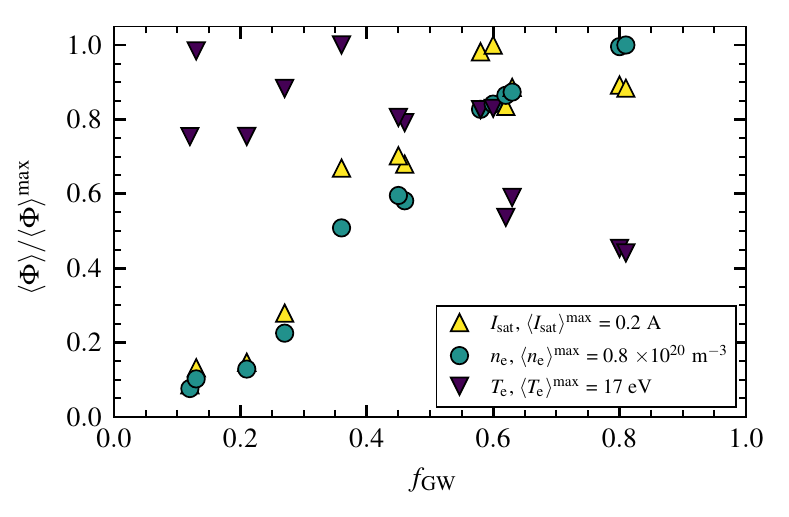}\label{fig:asp-mlp-mean}}
    \subfigure[]{\includegraphics[width=0.48\textwidth]{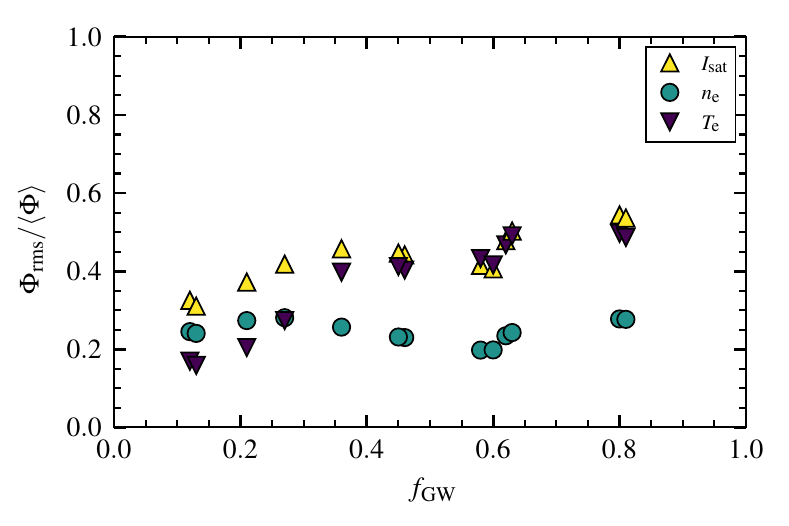}\label{fig:asp-mlp-rfl}}\\
    \caption{Plasma parameters (shown in the legend with the maximal value across all Greenwald fractions), as a function of Greenwald fraction from the dwell MLP density scan, where (a) shows the normalized mean value of the plasma parameter and (b) shows the relative fluctuation levels of the plasma parameter for the length of the time windows shown in table \ref{tab:asp-mlp-density-table}.}
    \label{fig:asp-mlp-mean-rfl-plasma-parameters}
\end{figure*}

The mean plasma parameters in the far-SOL versus Greenwald fraction, as measured by the MLP in dwell mode, are shown on the left-hand side of figure \ref{fig:asp-mlp-mean-rfl-plasma-parameters}. In figure \ref{fig:asp-mlp-mean}, $\avg{\isat}$ and $\avg{\nee}$ increase with the Greenwald fraction, but saturate at high densities where $\fg > 0.6$.  $\avg{\Te}$ shows the opposite dependence, decreasing with the line-average density. This is expected if the power flow to the far-SOL does not increase in the same proportion as the density increase. The relative fluctuations estimated as a ratio between the root mean square of the plasma parameter and the mean plasma parameter are shown in figure \ref{fig:asp-mlp-rfl}. The $\isat$ relative fluctuations increase by a factor of two while the $\nee$ relative fluctuations are roughly constant. Although $\avg{\Te}$ decreases with Greenwald fraction, the relative fluctuation levels in $\Te$ increase with density by a factor of more than two.

\begin{figure*}
    \centering
    \subfigure[]{\includegraphics[width=0.48\textwidth]{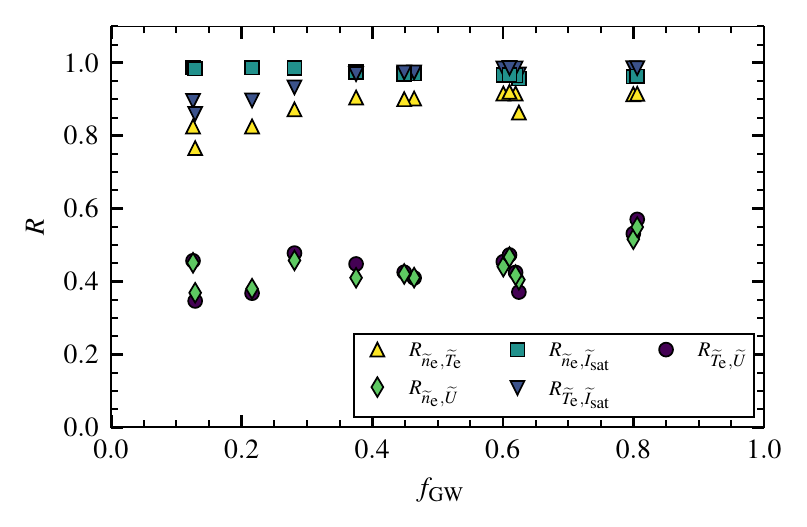}\label{fig:asp-mlp-correlations-ne-Te-U}}
     \subfigure[]{\includegraphics[width=0.48\textwidth]{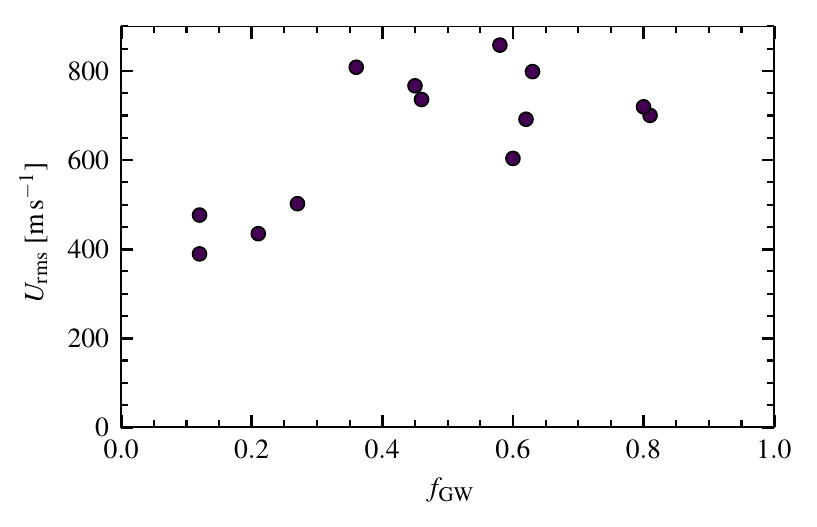}\label{fig:asp-mlp-correlations-U-rms}}
    \caption{Pearson correlation coefficients $R$ of (a) the normalized electron density fluctuations and the normalized electron temperature fluctuations $R_{\wt{n}_\mathrm{e},\wt{T}_\mathrm{e}}$; the normalized electron density fluctuations and the normalized ion saturation current fluctuations $R_{\wt{n}_\mathrm{e},\wt{I}_\mathrm{sat}}$; the normalized electron temperature fluctuations and the normalized radial velocity fluctuations $R_{\wt{T}_\mathrm{e},\wt{U}}$; the normalized electron density fluctuations and the normalized radial velocity fluctuations $R_{\wt{T}_\mathrm{e},\wt{U}}$; the normalized electron temperature fluctuations and the normalized ion saturation current fluctuations $R_{\wt{T}_\mathrm{e},\wt{I}_\mathrm{sat}}$ and (b) the root mean square of the radial velocity fluctuations $U_\mathrm{rms}$ as a function of Greenwald fraction from the density scan.}
    \label{fig:asp-mlp-correlations-vs-fgw}
\end{figure*}

The Pearson correlation coefficients $R$ of the plasma parameter fluctuations are shown in figure \ref{fig:asp-mlp-correlations-ne-Te-U}. Once again, these are from the MLP in dwell mode. For the reader's interest, the joint PDFs of these plasma discharges using the MLP from this runday have already been shown in a previous study \cite{kube-2019-nme}. The $\wt{n}_\mathrm{e}$, $\wt{T}_\mathrm{e}$ and $\wt{I}_\mathrm{sat}$ fluctuations are seen to be strongly correlated for all Greenwald fractions, and the correlation coefficient is practically independent of Greenwald fraction. Normalized radial velocity fluctuations $\wt{U}$ were calculated using the plasma potential as described by \eqref{eq:radial-velocity}. Fluctuations in particle density and electron temperature are both positively correlated with fluctuations in the radial velocity, and the correlation coefficients are practicaly identical and independent of Greenwald fraction. These correlations are not as strong as the correlation between particle density and temperature. While the correlations do not change significantly with density, the root mean square of the radial velocity fluctuations $\rms{U}$, shows a weakly increasing dependence, as shown in figure \ref{fig:asp-mlp-correlations-U-rms}. By relating the average size of the velocity to $\taud / U_\mathrm{rms}$, this suggests that on average, the sizes of these filaments are also weakly increasing since $\taud$ remains constant. The radial velocity of fluctuations increasing with line-average density have been reported previously in Alcator C-Mod \cite{agostini-2011,kube-2013}. Considered together, these correlations suggest that large fluctuations in particle density are associated with large fluctuations in electron temperature as well as positive radial velocities. 

\begin{figure*}
    \centering
    \subfigure[]{\includegraphics[width=0.48\textwidth]{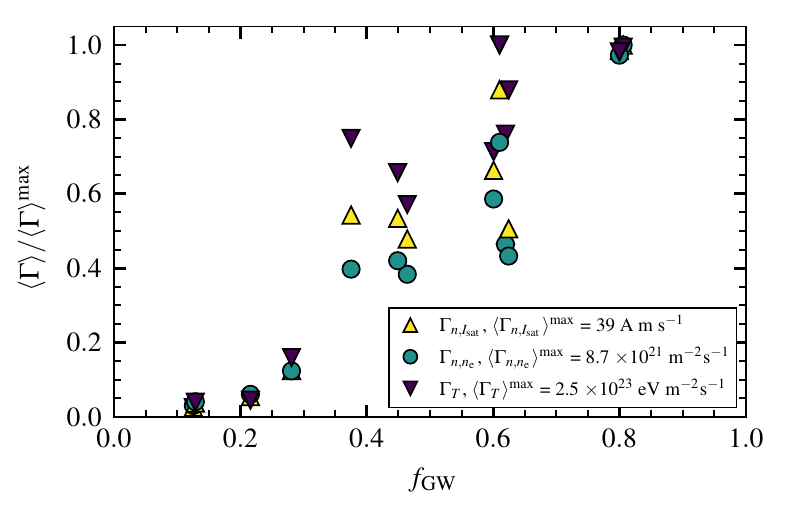}\label{fig:flux-all-vs-fgw}}
    \subfigure[]{\includegraphics[width=0.48\textwidth]{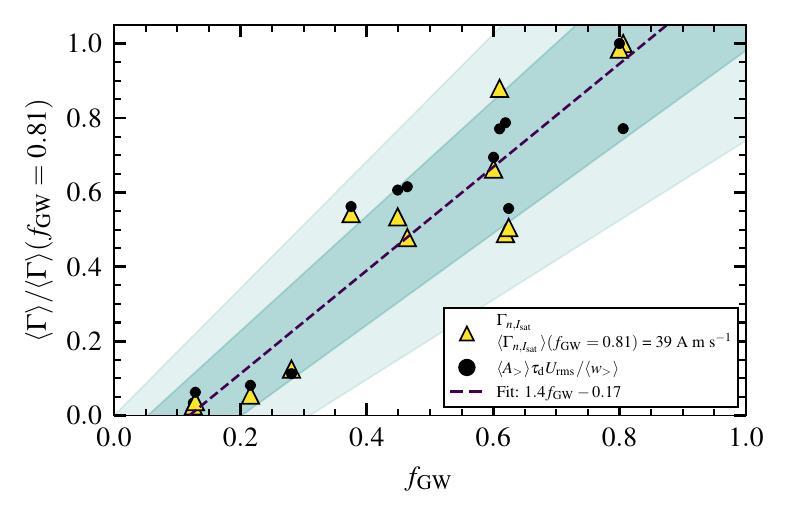}\label{fig:flux-scaling-vs-fgw}}
    \caption{The radial fluxes measures for varying Greenwald fraction from the density scan where (a) shows the mean radial particle flux using $\nee$ fluctuation data, $\isat$ fluctuation data and the mean heat flux and (b) shows the how the flux scales with $\avg{A_>}\taud U_\mathrm{rms}/\avg{w_>}$ (black circles) where $\avg{A_>}$ and $\avg{w_>}$ are estimated from the deconvolution (see figure \ref{fig:fpp-vs-fgw}) and $\taud$ is estimated from the PSD. The dashed line is the linear fit to the statistical estimates of the flux. The shaded region shows the confidence intervals of $1\sigma$ and $2\sigma$ from the linear fit.}
    \label{fig:asp-mlp-radial-particle-heat-flux}
\end{figure*}

We explore the radial particle flux using the $\nee$ and $\isat$ fluctuation measurements and the total heat flux and how this scales using the stochastic model parameters in figure \ref{fig:asp-mlp-radial-particle-heat-flux} as a function of the Greenwald fraction from the density scan. We present the mean values of the particle density flux using the electron density measurements $\avg{\Gamma_{n,\nee}}$, the mean values of particle flux using the ion saturation current measurements $\avg{\Gamma_{n,\isat}}$ and the mean heat flux $\avg{\Gamma_{T}}$ which are calculated as
\begin{equation}
\avg{\Gamma_{n, n_\text{e}}} = n_\text{e,rms}U_\mathrm{rms}\avg{\wt{n}_\text{e}\wt{U}},
\end{equation}
\begin{equation}
\avg{\Gamma_{n, I_\text{sat}}} = I_\text{sat,rms}U_\mathrm{rms}\avg{\wt{I}_\text{sat}\wt{U}},
\end{equation}
\begin{equation}
\avg{\Gamma_T} = U_\mathrm{rms} \biggl \langle \wt{U}\left[\Ttm{\Te}\wt{n}_\text{e}n_\text{e,rms} + \Ttm{\nee}\wt{T}_\text{e}T_\text{e,rms} + \wt{n}_\text{e}n_\text{e,rms} \wt{T}_\text{e}T_\text{e,rms}\right]\biggr \rangle.
\end{equation}

In figure \ref{fig:flux-all-vs-fgw}, as expected, the radial particle flux increases with line-averaged density, indicating more transport where the filaments are hotter and larger in amplitude. The mean of the total heat flux, as calculated from \eqref{eq:heat-flux}, increases with density. Figure \ref{fig:flux-scaling-vs-fgw} shows the scaling of $\avg{\Gamma_{n,\isat}}$  calculated using $\avg{\Gamma_{n,\isat}}\sim \Ttm{A}\taud  U_\mathrm{rms}/\Ttm{w}$ since this combination of statistical quantities should be approximately proportional to the particle flux resulting from the radial motion of the filaments. We compare estimates of $\avg{A_>}\taud U_\mathrm{rms}/\avg{w_>}$ to $\avg{\Gamma_{n, \isat}}$ as we have estimated the mean amplitudes and mean waiting times from the deconvolution of the $\Isat$ and estimate $\taud$ from the PSD of the $\Isat$ measurements as shown in figure \ref{fig:density-scan-psd-sig-mlp}. The strong similarity in the scaling of these independently arrived at quantities serves as a consistency check and to increase confidence in the measurements. The minimum mean flux measurements were found to be all at $\fg=0.13$ where this was $\avg{\Gamma_{n, n_\text{e}}} =  2.3 \times 10^{20}\,\text{m}^{-2}\text{s}^{-1}$, $\avg{\Gamma_{n, I_\text{sat}}} = 1.0 \,\text{A}\,\text{m}\,\text{s}^{-1}$ and $\avg{\Gamma_T} = 6.3 \times 10^{21}\,\text{eV}\,\text{m}^{-2}\text{s}^{-1}$; to avoid confusion that these markers in figure \ref{fig:flux-all-vs-fgw} appear to be at zero. These are all approximately $1/40$ times their respective maximal values.

To summarize figures \ref{fig:asp-mlp-mean-rfl-plasma-parameters}, \ref{fig:asp-mlp-correlations-vs-fgw}, and \ref{fig:asp-mlp-radial-particle-heat-flux}, the largest filaments are hot, dense and fast compared to the background values, and their velocities appear to be increasing with Greenwald fraction. Although $\taud$ remains approximately constant, increasing velocity indicates that the filaments are becoming larger in size. The independence of the correlations on the Greenwald fraction suggests that the physical mechanism of the blobs in the far-SOL is robust with Greenwald fraction. 

\begin{figure*}
    \centering
    \subfigure[]{\includegraphics[width=0.48\textwidth]{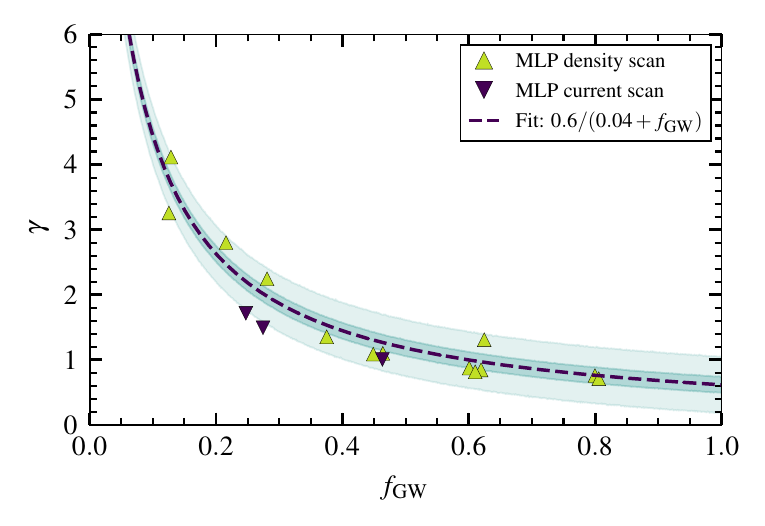}\label{fig:mlp-gamma-vs-fgw}}
    \subfigure[]{\includegraphics[width=0.48\textwidth]{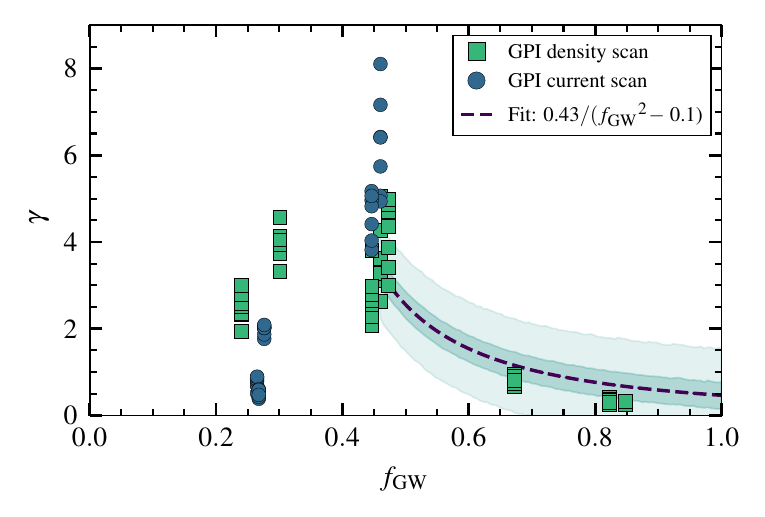}\label{fig:gpi-gamma-vs-fgw}}\\
    \subfigure[]{\includegraphics[width=0.48\textwidth]{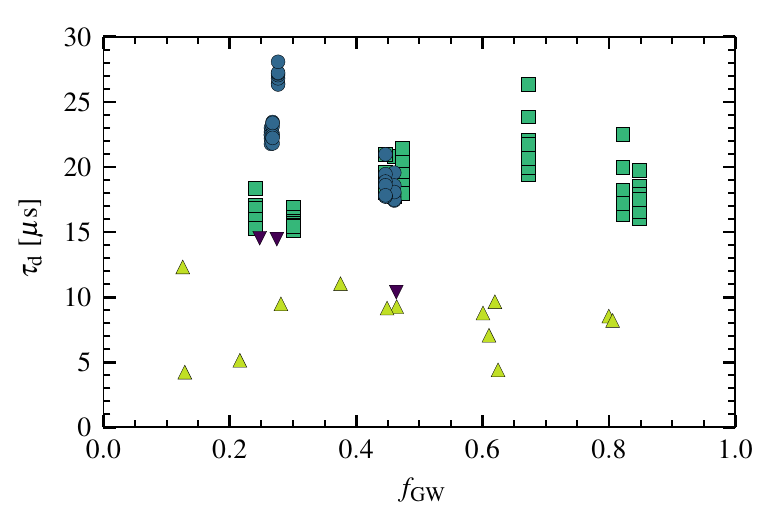}\label{fig:td-vs-fgw}}
    \subfigure[]{\includegraphics[width=0.48\textwidth]{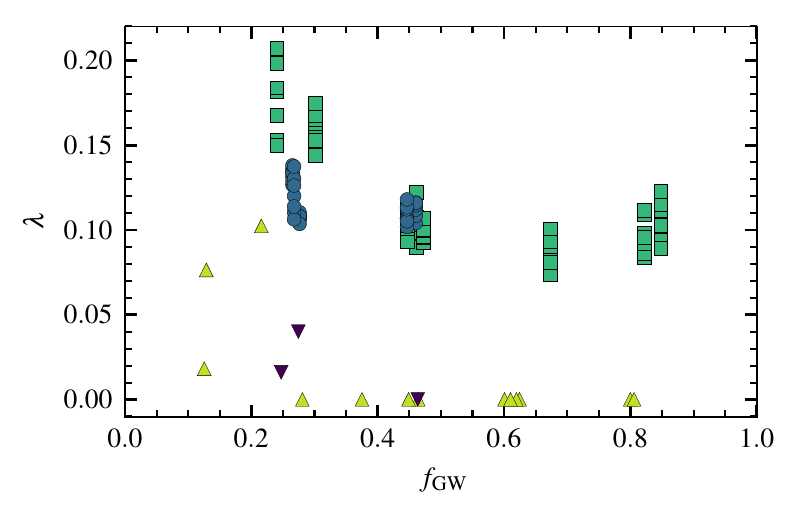}\label{fig:lam-vs-fgw}}\\
    \subfigure[]{\includegraphics[width=0.48 \textwidth]{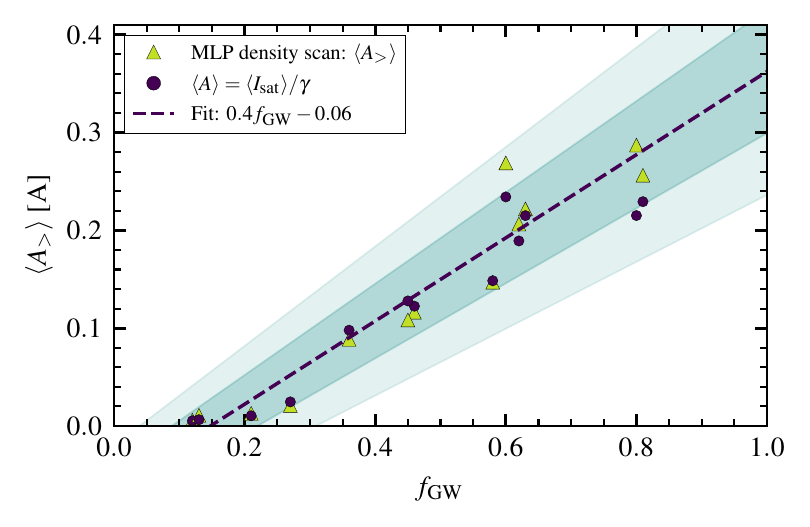}\label{fig:asp-mlp-density-mean-amp-Is}}
    \subfigure[]{\includegraphics[width=0.48\textwidth]{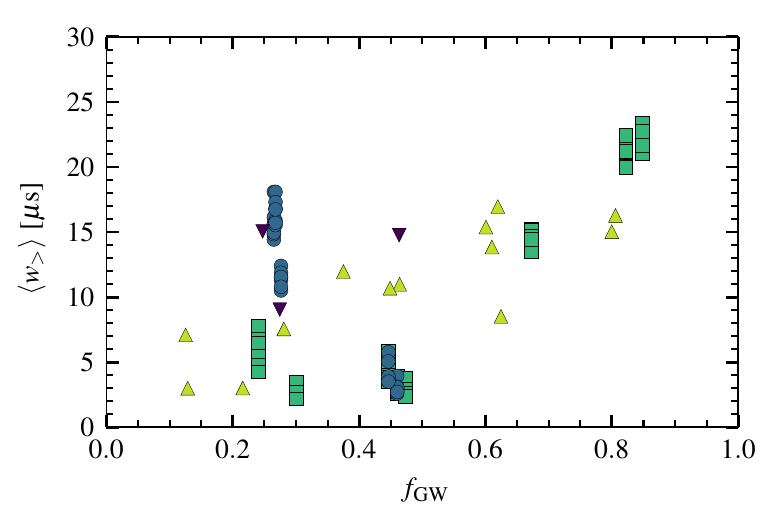}\label{fig:mean-wait-vs-fgw}}
    \caption{Summary of figures showing how the stochastic model parameters, estimated from the density scan and plasma current scan, change with Greenwald fraction: (a) Intermittency parameter from the MLP density and current scan where we employ an asymptotic fit (purple dashed-line); (b) Intermittency parameter from the GPI density and current scan showing a reciprocal square fit (purple dashed-line) for $\fg> 0.45$; (c) the duration time; (d) asymmetry parameter; (e) estimated mean amplitude $\avg{A_>}$ shown are only from the MLP $\Isat$ density scan as a function of Greenwald fraction. These are compared to $\Ttm{A}=\avg{\isat}/\gamma$ (purple circles) where we show the linear fit (purple dashed-line). (f) shows the estimated mean waiting time $\avg{w_>}$, from the bi-exponential fit. The markers represent which diagnostic and the type of scan in which the model parameters are estimated which can be inferred to the legends in (a) and (b). The shaded regions show confidence intervals of $1\sigma$ and $2\sigma$.} 
    \label{fig:fpp-vs-fgw}
\end{figure*}

The stochastic model parameters estimated from the line-averaged density and plasma current scans are presented in figure \ref{fig:fpp-vs-fgw} and are plotted against Greenwald fraction. The intermittency parameter $\gamma$ is shown in figures \ref{fig:mlp-gamma-vs-fgw} and \ref{fig:gpi-gamma-vs-fgw} for the MLP and GPI, respectively. We also show the average duration time $\taud$ in figure \ref{fig:td-vs-fgw}, the mean of the estimated waiting time $\avg{w_>}$ in figure \ref{fig:mean-wait-vs-fgw} and the mean of the estimated amplitudes $\avg{A_>}$ in figure \ref{fig:asp-mlp-density-mean-amp-Is}. The mean amplitudes are dimensionalized by $\avg{\isat}/\gamma$. The legend in figures \ref{fig:mlp-gamma-vs-fgw} and \ref{fig:gpi-gamma-vs-fgw} refers to the estimated parameters from the type of diagnostic and parameter scan. For the reader's interest, results showing how the fluctuation statistics change with respect to $\Ip$ instead of Greenwald fraction can be found in \ref{appendix:ip-scan}.

Firstly, it is clearly shown from the line-averaged density scan that as the core density is increased, the fluctuations become increasingly intermittent for both the GPI (green squares) and the MLP (yellow triangles) for $\fg \gtrsim 0.45$ as shown in figures \ref{fig:mlp-gamma-vs-fgw} and \ref{fig:gpi-gamma-vs-fgw}. The MLP shows an asymptotic decrease in the value of $\gamma$. A parabolic fit predicts an increase which is not seen in the data, therefore we employ an asymptotic regression on the MLP $\gamma$ results revealing a scaling of $\sim 1 / \fg$. The MLP plasma current scan (purple inverted triangles) also shows that the fluctuations become strongly intermittent with increasing Greenwald fraction (decreasing $\Ip$) but this dependence appears somewhat weaker than that resulting from an increase in density. Nonetheless, this result aligns well with the line-averaged density scan using the MLP. 

However, the intermittency parameter estimated from the GPI plasma current scan (blue circles) increases from $\fg=0.27$ to $\fg=0.46$. This increase in the intermittency parameter (where fluctuations are becoming weakly intermittent) between low-intermediate Greenwald fractions has also been seen in previous results \cite{garcia-2013}. The GPI measurements from the density scan and plasma current scan show a spread in the estimated parameters. The diode views considered are nominally from similar flux positions and show some variation in the PDFs. These scans show similar trends where the intermittency increases with Greenwald fraction (that is, fluctuations becoming weakly intermittent). We utilize an asymptotic regression for the GPI $\gamma$ results in figure \ref{fig:gpi-gamma-vs-fgw} from $\fg \sim 0.46$ to describe the decrease in $\gamma$, which revealed a scaling of $\sim 1 / \fg^2$.

There is a spread in the intermittency parameter estimated from the GPI measurements for intermediate Greenwald fractions around $\fg \sim 0.46$. Note that for shot 1160616018 ($\fg \sim 0.47$) the MLP and the GPI were operational, and it is striking to see the differences in the estimated intermittency parameter. For shots 1160616026 and 1160616027 ($\fg \sim 0.81$), the MLP and GPI were also operational but there are little differences in the intermittency parameters. This discrepancy in the intermittency parameter between the GPI and the MLP ion saturation current is also reported in \cite{kube-2020}, where the same discharge from Alcator C-Mod was analyzed. In \ref{appendix:fit-issues}, we investigate the quality of the fits to the GPI measurements from the mid-Greenwald fraction discharges $\fg = 0.47$ and found that these GPI measurements do not have an elevated tail and show poor agreement with the stochastic model. Thus the fitted values of $\gamma$ depend heavily on the fit ranges, explaining some of the spread in the results. In contrast, the distributions of the GPI normalized time series for $\fg > 0.6$ are well fitted. The MLP $\Isat$ measurements show good agreement for all Greenwald fractions.

Next, in figure \ref{fig:td-vs-fgw}, the duration times estimated from the far-SOL GPI and MLP measurements seem robust against changes in line-averaged density. On average, the duration times estimated from the GPI measurements are larger than those found the MLP measurements. The small duration times seen from some of the MLP measurements may be due to strong poloidal velocities present in the SOL as also discussed in reference \cite{agostini-2011}. The larger estimates of the pulse duration times from the GPI is due, at least in part, to spatial averaging. For example, a $1\,\text{cm}$ filament moving past a probe tip at $1\,\text{km}\,\text{s}^{-1}$, will have a $\taud \sim 10\,\mu\text{s}$. If a single GPI view averages over a  $0.7\,\text{cm}$ radial region, then it will measure a $\taud \sim 17\,\mu\text{s}$ for the same $1\,\text{cm}, 1\,\text{km}\,\text{s}^{-1}$ filament. A $0.7\,\text{cm}$ spatial smoothing, while larger than the $0.38\,\text{cm}$ optical in-focus spot size, can easily result from the finite size of the gas cloud and the $\sim 8^{\circ}$ angle of the viewing chord relative to the local fieldline and provide a likely reason for the difference in the estimated $\taud$ values measured by the two diagnostics. The effects of spatial averaging would also impact the estimated $\lambda$, resulting in larger values, and hence a more symmetric pulse shape compared to the MLP. The observation that the discrepancies in the GPI and MLP evaluations of $\taud$ and $\lambda$ are likely due to the poorer GPI spatial resolution is not surprising since, as demonstrated, spatial resolution can enter into the evaluations if it is not significantly smaller than the typical blob-size. It is therefore important that this effect be considered when performing detailed fluctuation analyses. However, it does not render the finite-resolution GPI measurements unsuitable for such analyses since GPI measurements still provide valuable long time-series data over the GPI field-of-view and under edge plasma conditions where heat-fluxes are too large for a scanning probe. 

Although the plasma current scan shows far fewer data points than the density scan, there seems to be no significant trend in $\taud$ with Greenwald fraction. The MLP current scan points are within the scatter of the MLP density scan. Once again, the duration times estimated from the GPI plasma current scan are larger compared to those estimated from the MLP plasma current scan.

The asymmetry parameter for the various scans is shown in figure \ref{fig:lam-vs-fgw}. $\lambda$ values between 0.02 and 0.1 are measured by the MLP for $\fg < 0.3$. The very small $\lambda$ values estimated at higher Greenwald fractions are due to reaching the lower limits of the fitting technique, suggesting a pulse shape that is close to a one-side exponential with a very fast initial rise. The $\lambda$ evaluations from the GPI data show a decrease in $\lambda$ as $\fg$ increases to $\sim0.45$. Above that Greenwald fraction, $\lambda$ is essentially constant. This suggests that the shape of the fluctuations, on average, appear more asymmteric at higher densities. The results from the GPI plasma current results are within the scatter of the GPI density scan.  

In figure \ref{fig:asp-mlp-density-mean-amp-Is}, the mean amplitudes of the ion saturation current fluctuations are presented in physical units for all densities considered. These mean values were achieved by performing bi-exponential fits to the estimated amplitude distribution using \eqref{eq:bi-exp-amp} to get $\Ttm{A_>}$ which is then multiplied by $\avg{\isat}/\gamma$. Since we are interested in the large-amplitude fluctuations under the reasonable assumption that they dominate the cross-field transport, we present the estimated mean amplitude of these fluctuations. As expected, the mean amplitudes of the fluctuations increase with the core density, indicating that these strongly intermittent fluctuations become increasingly large. In conjunction with the flux of the particle density shown previously in figure \ref{fig:asp-mlp-radial-particle-heat-flux}, at high Greenwald fractions there is a higher level of particle transport with fluctuations in the far-SOL, driven by large-amplitude events. 

\begin{figure*}
    \centering
\subfigure[]{\includegraphics[width=0.48\textwidth]{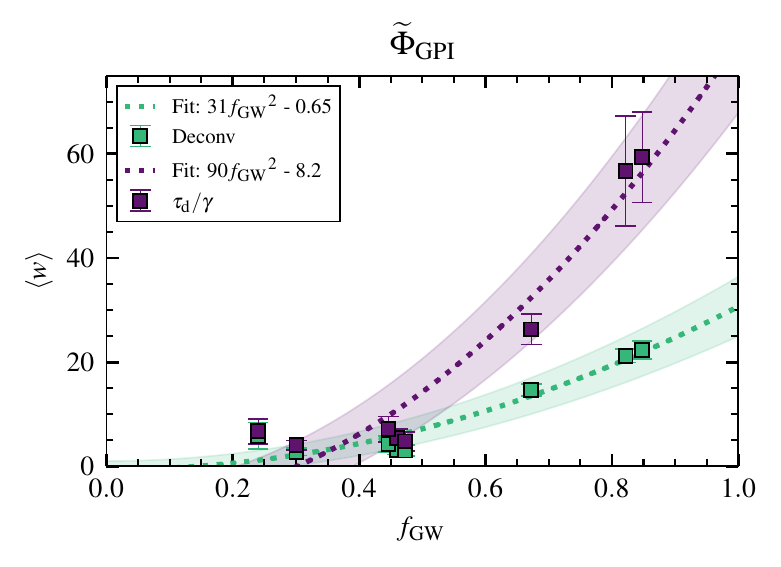}\label{fig:gpi-deconv-stats}}
\subfigure[]{\includegraphics[width=0.48\textwidth]{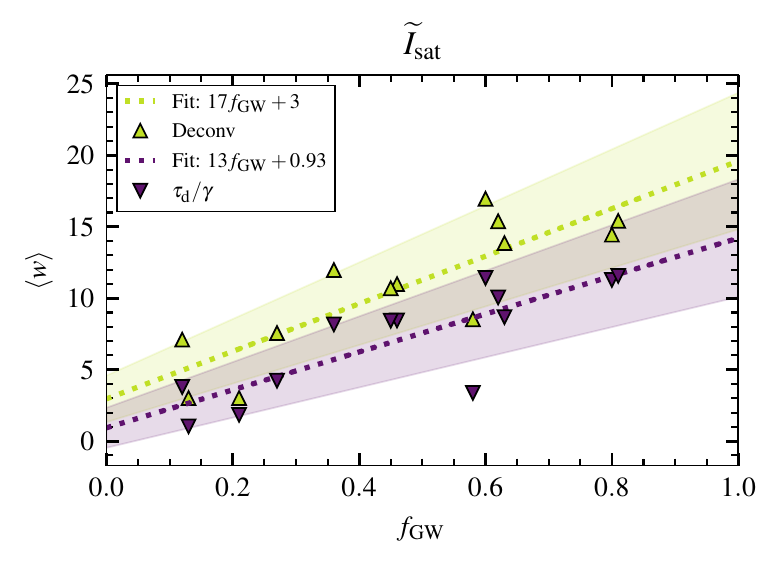}\label{fig:mlp-deconv-stats}}
    \caption{Comparison of the mean waiting times from the fluctuation statistics ($\taud/\gamma$) and the RL deconvolution algorithm represented by the markers. Focusing on results from the density scan, (a) GPI shows the average of the mean waiting times for all diode view positions considered where the error bars correspond to the maximum and minimum mean waiting times. The dotted lines show a non-linear regression for the GPI. (b) shows the mean waiting time estimated from the MLP $\Isat$ measurements. The dotted lines for the MLP scaling represent a linear regression. The shaded regions correspond to 1$\sigma$ intervals of the results.}
    \label{fig:deconv-stats-scaling}
\end{figure*}

The mean waiting times in figure \ref{fig:mean-wait-vs-fgw} are estimated by performing a bi-exponential fit to the deconvolved waiting time distribution to avoid the effects of noise and blob dissipation, which affect the smaller waiting times the most. We focus on the mean of the larger waiting times. In figure \ref{fig:mean-wait-vs-fgw}, the mean waiting times from the two diagnostics are roughly similar at similar Greenwald fractions for both scans, except at $\fg$=0.3 and 0.45, where the GPI-measured waiting times are smaller than those measured by the MLP. The reason for this is not known and is confusing since we expect that the GPI-measured waiting times should be greater than or equal to those from the MLP since the GPI measures at $\rho$ values larger than those for the MLP, as will be discussed below. Nevertheless, the overall trend with Greenwald fraction is similar and notable. 

We look at a comparison of the mean waiting times in figure \ref{fig:deconv-stats-scaling} between the estimation from the fluctuation statistics (i.e. $\Ttm{w}=\taud/\gamma$, where $\taud$ is evaluated from the PSD and $\gamma$ is evaluated from the ECF) and from the deconvolution algorithm. The methods used to estimate $\gamma$ and $\taud$ involve no thresholding which will lead to shorter mean waiting times. For the GPI mean waiting times, we employ a non-linear regression $\avg{w} =  \alpha \fg^{2} + c_1$ , where the slope of this regression is $\alpha$ and $c_1$ is the intercept. In contrast, a linear regression $\avg{w} =  \beta \fg + c_2$, was performed on the MLP mean waiting times where the slope of this scaling is referred to and $\beta$ and the intercept it given by $c_2$. 

We present these scalings for the GPI in figure \ref{fig:gpi-deconv-stats} showing the mean waiting times estimated from the deconvolution algorithm and the mean waiting times estimated from the fluctuation statistics. We see a divergence of these results for $\fg>0.5$ where these mean waiting times are unequal. The regression reveals a scaling of $\alpha_{\taud/\gamma}$ = 90 for $\Ttm{w} = \taud/\gamma$ and $\alpha_\mathrm{Deconv}$ = 31 for $\avg{w_>}$ estimated from the bi-exponential fit to the deconvolved waiting times. We will present the MLP results before discussing this discrepancy.  

For the MLP $\Isat$ measurements, shown in figure \ref{fig:mlp-deconv-stats}, the regression reveals a scaling of $\beta_{\taud/\gamma}$ = 13 for $\Ttm{w} = \taud/\gamma$ and $\beta_\mathrm{Deconv}$ = 17 for $\avg{w_>}$ estimated from the bi-exponential fit to the deconvolved waiting times. However, the mean waiting times calculated from $\taud/\gamma$ are lower than the mean waiting time estimated from the bi-exponential fit on the deconvolved waiting time distribution. This is due to the threshold-independent estimation of $\gamma$ and $\taud$ taking into account all of the fluctuations in the time series, which will naturally lead to slightly lower mean waiting times.

We now turn to the discrepancy between $\Ttm{w}$ and $\avg{w_>}$ seen for the GPI but not for the MLP in figure \ref{fig:deconv-stats-scaling}. For the MLP, figure \ref{fig:density-scan-pdf-amp-wait-mlp} shows that the large amplitudes and waiting times indeed follow exponential distributions consistent with the FPP. That is corroberated by the favorable comparison in figure \ref{fig:mlp-deconv-stats} and indicates that the small amplitudes and waiting times are mainly due to the noise process. For the GPI, figure \ref{fig:density-scan-pdf-amp-wait-gpi} shows the same exponentially distributed large waiting times, but we did not produce convincing fits to the amplitude distribution in figure \ref{fig:gpi-density-pdf-amp}. Using the non-exponential amplitude distribution of the GPI as a working hypothesis, the discrepancy between $\Ttm{w}$ and $\avg{w_>}$ for the GPI may be explained as follows: Numerical testing suggests that a positive definite amplitude distribution with higher flatness than an exponential distribution leads us to underestimate $\gamma$, if we make a fit using the Gamma distribution from the standard FPP with exponential amplitudes. If this is the case, as it is for the amplitudes in figure \ref{fig:gpi-density-pdf-amp}, $\gamma$ has been underestimated for the GPI  leading to an overestimate of $\Ttm{w} = \taud/\gamma$, consistent with figure \ref{fig:gpi-deconv-stats}. The opposite effect, that a positive definite amplitude distribution with lower flatness than the exponential distribution leads us to underestimate $\gamma$, is seen in figure 3 in reference \cite{theodorsen-2018-ppcf}. This explanation remains tentative, however, as a wrong estimate of $\gamma$ leads to an incorrect rescaling $\sqrt{\gamma (1+\eps)} \widetilde{\Phi} + \gamma$, it also influences the deconvolution and in turn the deconvolved amplitude distribution. A consistent estimate requires further modelling work.

\begin{figure*}
    \centering
    \subfigure[]{\includegraphics[width=0.48\textwidth]{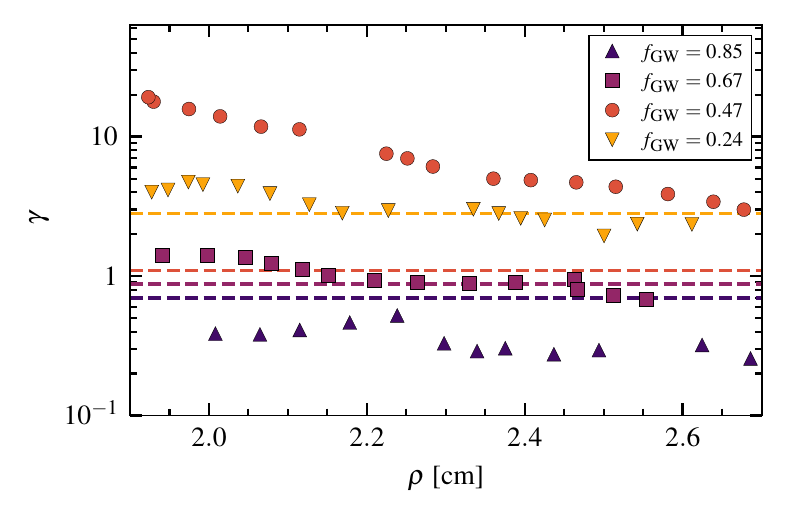}\label{fig:gpi-gamma-vs-rho}}
    \subfigure[]{\includegraphics[width=0.48\textwidth]{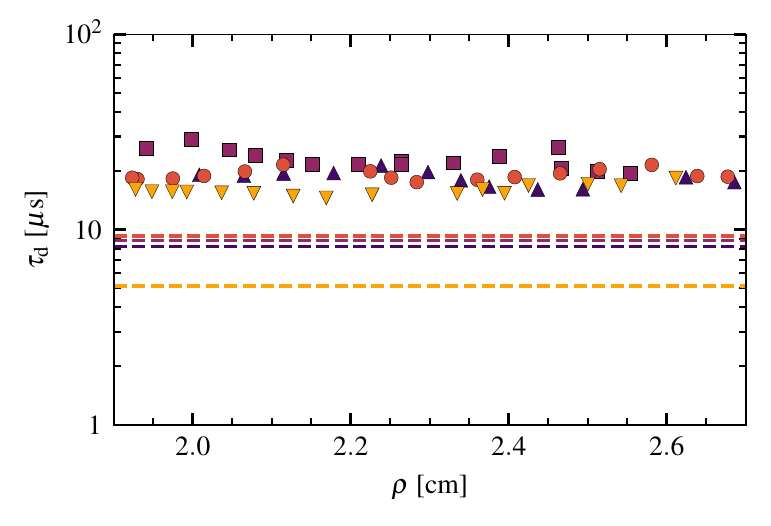}\label{fig:gpi-td-vs-rho}}\\
     \subfigure[]{\includegraphics[width=0.48\textwidth]{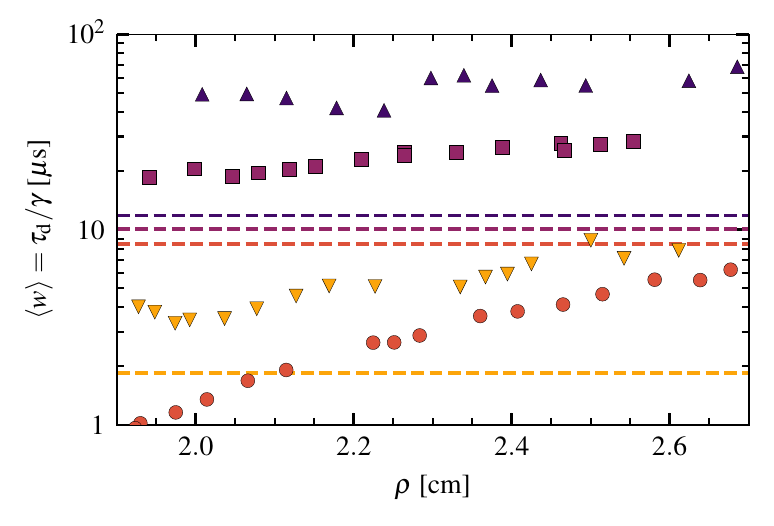}\label{fig:gpi-mean-wait-vs-rho}}
    \caption{The stochastic model parameters estimated from the GPI square markers) where (a) $\gamma$, (b) $\taud$ and (c) $\Ttm{w} = \taud/\gamma$ are plotted as a function of the GPI view location in terms of $\rho$. Here, we take into consideration a larger range of views compared to what is shown in previous figures to demonstrate how the statistics change with $\rho$. The legend is shown in (a) where the colors represent the Greenwald fractions from the GPI density scan. The MLP results are plotted as dashed lines at similar Greenwald fractions to the GPI results, indicated by the same colors as shown in the legend. The MLP measures the fluctuations at a significantly smaller $\rho$ ($\sim$ 0.9-1.4 cm, see table \ref{tab:asp-mlp-density-table}) compared to the GPI, but is shown here for comparison.}
    \label{fig:gpi-fpp-vs-rho}
\end{figure*}

Investigating the GPI measurements further, we present how $\rho$ may influence the statistics at various Greenwald fraction densities. This is presented in figure \ref{fig:gpi-fpp-vs-rho} where we also show the MLP results at similar densities to the GPI as colored dashed lines for comparison and we recognize that these fluctuations measured by the MLP were taken at a lower $\rho$ location, as seen in table \ref{tab:asp-mlp-density-table}. $\gamma$ as a function of $\rho$ is presented in figure \ref{fig:gpi-gamma-vs-rho} for all of the diode views considered in this study. For $\fg \leq 0.47$, the value of $\gamma$ decreases (becomes more intermittent) as $\rho$ increases which is consistent with previous results \cite{theodorsen-2017-nf}. The other Greenwald fraction cases are decreasing in the ranges of $\rho$ shown in figure \ref{fig:gpi-gamma-vs-rho} but not as strongly as $\fg=0.47$. As $\rho$ increases, filaments experience dissipation through parallel transport as they propagate through the SOL. As we approach higher line-averaged densities where $\fg \geq 0.67$, we have strongly intermittent fluctuations where $\gamma < 1$ across the $\rho$-space presented in figure \ref{fig:gpi-gamma-vs-rho} \cite{garcia-2016}. This is due to the flattening and broadening of the SOL profiles observed in figure \ref{fig:asp-mlp-density-profiles}. The range of MLP $\gamma$ values shows that these are smaller compared to the GPI estimate at $\fg=0.47$. However, for the highest density case, variation in $\rho$ makes little difference to the intermittency values from both diagnostics.

The pulse duration times as a function of $\rho$ in figure \ref{fig:gpi-td-vs-rho} seem to vary little, consistent with previous results \cite{theodorsen-2017-nf}. We have explained earlier the differences in the estimation of $\taud$ between the MLP and GPI. In figure \ref{fig:gpi-mean-wait-vs-rho} where $\Ttm{w} = \taud / \gamma$, these means increase with $\rho$ for the Greenwald fractions presented here which is indicative of the effects of pulse overlap shown in figure \ref{fig:gpi-gamma-vs-rho}. In particular, $\fg=0.85$ shows that the values for $\Ttm{w}$ remain constant with $\rho$ due to flat profiles, as previously discussed. For $\fg \geq 0.67$, $\Ttm{w}$ from the MLP are lower overall compared to the high density GPI results. The GPI views are seeing fluctuations at a larger $\rho$ location, hence larger $\Ttm{w}$.

In general, figure \ref{fig:fpp-vs-fgw} summarizes the parametric analysis when it comes to changing the line-averaged density and plasma current in Alcator C-Mod. From both diagnostics, we see the same trends with fluctuations becoming strongly intermittent, increasing mean waiting times, and no change in the average duration times with increasing Greenwald fraction. At some values of $\fg$, the actual values $\avg{w_>}$ and $\gamma$ are similar, while at others they disagree but within a factor of two. The actual values of $\taud$ differ consistently by a factor of approximately two.

\section{Discussion and conclusions}\label{sec:discussion-conclusions}

The fluctuation statistics for various plasma parameters in ohmic, diverted single-null configurations are presented, using time-series data from GPI and the MLP. The profiles of the relative fluctuation levels with increasing main-plasma line-averaged density suggest that fluctuations become more intermittent. As the densities increase, the observed shift in shape of the histograms obtained from the normalized time series can be attributed to the reduced occurrence of filaments with higher mean amplitudes and velocities. The agreement with the stochastic model at low line-averaged densities is well established \cite{garcia-2013,garcia-2018,theodorsen-2017-nf,theodorsen-2018-php}. Here we demonstrate that this agreement continues to hold for high line-averaged densities, indicating that the mechanism behind generating these filaments does not change with density.

The deconvolution algorithm was used to recover the pulse amplitudes and arrival times, as opposed to the conditional averaging technique, in order to provide statistics for a much more inclusive range of fluctuation amplitudes.  Large filaments are believed to have an outsized contribution to plasma--wall interactions. Therefore, we only consider the large amplitudes and large waiting times in the analysis of the deconvolution. The deconvolution shows that the mean amplitudes and the mean waiting times increase for the ion saturation current measurements at increasing line-average densities. Increasing mean waiting times is contrary to the idea that the filaments occur more frequently as the density is increased which was an observation made from GPI measurements at low densities \cite{garcia-2013,garcia-2018,theodorsen-2017-nf}. However, our analysis shows that this is not the case for the MLP. Despite the fluctuations occuring less frequently at high densities, the particle and heat fluxes increase significantly where the filament amplitudes are considerably large, thereby amplifying plasma-wall interactions.

The shape of normalized frequency power spectra appears independent of the Greenwald fraction. This means that the temporal scale of these fluctuations, as parameterized by the quantity, $\taud$, stays roughly constant as the density increases. Inferring from the fluctuation statistics at lower Greenwald fractions, signals from both the GPI and the MLP show more pulse overlap compared to higher Greenwald fraction cases where the pulses appear more isolated and larger in amplitude. 

Differences in duration times and mean amplitudes between GPI and MLP can be attributed to the effects of spatial averaging as explained in the previous section. The factor of two difference between the GPI and MLP estimated duration times are reasonable. The MLP shows shorter duration times estimated from the frequency power spectral density of the far-SOL measurements with no dependence on the Greenwald fraction. Furthermore, absolute values of the amplitudes can be recovered from the MLP measurements.

In references \cite{garcia-2013} and \cite{garcia-2018} it was shown that $\gamma$ increases with line-averaged density up to $\fg=0.35$. Observing these time series from the outermost diode view only, reference \cite{garcia-2013} used a four-point density scan up to $\fg=0.35$ and showed the same trends as shown in figure \ref{fig:gpi-gamma-vs-fgw}. There was not enough from the two-point density scan in reference \cite{garcia-2018} to see a clear trend. These studies did not go to high enough Greenwald fractions in density to see that the value of the intermittency parameter decreases with Greenwald fraction. It remains unclear as to why the $\gamma$ value peaks at $\fg \sim 0.46$ for the GPI where these observations are not seen in the MLP $\gamma$ scaling with Greenwald fraction. 

There are strong correlations between the $\wt{n}_\mathrm{e}$, $\wt{U}$ and $\wt{T}_\mathrm{e}$ fluctuations. These correlations are constant throughout the line-average density scan, indicating that the physics behind filament propagation at increasing line-averaged densities is not changing. However, the $U_\mathrm{rms}$ increases with the line-average density, suggesting that the size of the filaments is increasing since the durations times are constant. Consequently, at higher line-average densities, intermittent and large-amplitude fluctuations of the electron density and temperatures at high velocities will increase plasma-wall interaction. Density and velocity fluctuations that appear in phase, as demonstrated by the correlation study, lead to large particle flux events. This is consistent with the observation that the midplane neutral pressure increases with line-average density \cite{labombard-2000}.

\subsection{Comparison to other devices}

Previously, scans in plasma current were performed in TCV, MAST and DIII-D. In TCV, it was seen that the plasma current does not make a difference in the shape of the probability density functions of the Langmuir probe time series and the conditionally averaged waveforms \cite{garcia-2007-coll}. Also, at lower plasma currents, the mean profile becomes broader. Despite the change in collisionality in the study presented in reference \cite{garcia-2007-coll}, this feature did not make a difference in the shape of the probability density functions and hence the intermittency. In MAST, a plasma current scan revealed that the radial velocity and radial size of L-mode filaments decrease with increasing plasma current, thus decreasing radial transport \cite{kirk-2016}. In DIII-D, similar results were also observed in which plasma ion fluxes to the low-field side increased with decreasing plasma current \cite{rudakov-2005}. The radial mean profiles from the Alcator C-Mod plasma current analysis is in agreement with these previous findings.

Alcator C-Mod and TCV are most similar in machine size, but differ when it comes to divertor design \cite{Vianello_2020}. It was found in TCV that as the line-averaged density increases, the filament velocities become increasingly larger and the profiles become flatter and broader; features also exhibited in Alcator C-Mod \cite{offeddu_2022, garcia-2007-nf, kube-2013, kube-2019-nme}. Interestingly, it was found in TCV for various line-averaged densities in ohmic, diverted single-null plasmas that the PDFs of the probe measurements at the wall radius do not change as shown in reference \cite{garcia-2007-nf} which is in disagreement with the Alcator C-Mod density scan results.

The SPARC tokamak ($R=1.85\,\text{m}$ and $a=0.57\,\text{m}$) is planning to operate initially in L-mode \cite{creely2020}. Despite this study being a single-machine scan, we utilized the scalings presented in this study, and extrapolate to the outer limiters that SPARC will have a particle wall flux of $\sim 7 \times 10^{18} \,\text{m}^{-2}\text{s}^{-1}$. This is a decrease of approximately two orders of magnitude compared to Alcator C-Mod at similar Greenwald fractions.

\subsection{Conclusion}
Radially propagating far-SOL fluctuations become increasingly intermittent with Greenwald fraction. Notably, however, there exist significant differences between the results from GPI time-series (as measured at $\rho$ between $2.3$ and $2.7\,\text{cm}$) and MLP time-series (as measured at $\rho$ between $0.8$ and $1.4\,\text{cm}$) in terms of the average duration times across all Greenwald fractions, as well as the intermittency found at low-intermediate Greenwald fractions. These findings are intriguing and warrant further investigation in future research. Nonetheless, for $\fg \gtrsim 0.45$ it is worth noting that both diagnostics demonstrate strongly intermittent fluctuations with similar intermittency parameters. The continuous change in all statistical properties of the mirror-Langmuir probe data with Greenwald fraction indicates that there is no new physics mechanism as the empirical discharge density limit is approached.

We clarify the scalings of the stochastic model parameters with a wider range of Greenwald fractions which we have found for $\gamma$, $\avg{A}$, $\avg{w}$ and the particle flux. As a result of increasing line-averaged density, we see increasing temperatures of these fluctuations where filaments are getting hotter and move with increased radial velocities and amplitudes. Even though the mean waiting times between consecutive fluctuations get longer with increasing density, the filaments observed in the far-SOL are larger in amplitude, carrying radially most of the particle and heat. This leads to a significant increase in the plasma density at the wall. In turn, this will lead to increased plasma--wall interactions which will threaten the life-time of the first wall for future fusion devices and high duty cycle confinement experiments that plan to operate at high densities. Motivated by the scalings unraveled in order to inform predictive capability, we made some initial estimates for SPARC on the expected particle wall flux for an L-mode scenario.

Further work will focus on how the statistics on intermittent plasma fluctuations in the far-SOL change with machine size across various fusion devices, i.e. Alcator C-Mod, DIII-D, TCV and MAST. In addition, investigations are also underway on how profiles change with the Greenwald fraction, other plasma parameters, and confinement modes. This would use an extended version of the stochastic model recently developed by the UiT group that includes the radial position and the parallel drainage time \cite{losada-2022}.

\section*{Data availability statement}
The data cannot be made publicly available upon publication because no suitable repository exists for hosting data in this field of study. The data that support the findings of this study are available upon reasonable request from the authors.

\section*{Acknowledgments}
This work was supported by the Tromsø Research Foundation under grant number 19\_SG\_AT and the UiT Aurora Centre Program, UiT The Arctic University of Norway (2020). Alcator C-Mod data were generated and maintained under US Department of Energy awards DE-FC02-99ER54512 and DE-SC0014264. Support for A.Q.K.\ was provided by Commonwealth Fusion System under the grant number RPP-022. Support for J.L.T.\ was provided by the US Department of Energy, Fusion Energy Sciences, award DE-SC0014251. The first author thanks the MIT Plasma Science and Fusion Center for their generous hospitality where this work was conducted. Finally, discussions with Gregor Decristoforo are gratefully acknowledged.


\appendix

\section{Plasma current scan}\label{appendix:ip-scan}
\begin{figure}
    \centering
    \subfigure[]{\includegraphics[width=0.48\textwidth]{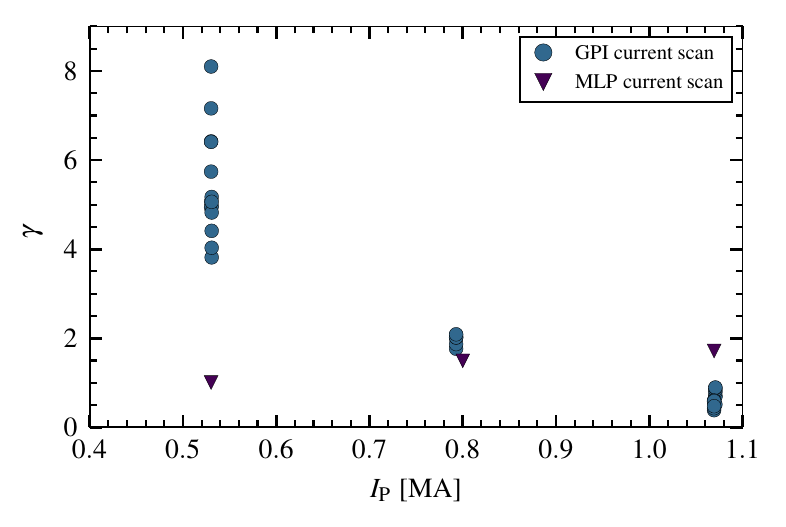}\label{fig:gamma-vs-current}}
    \subfigure[]{\includegraphics[width=0.48\textwidth]{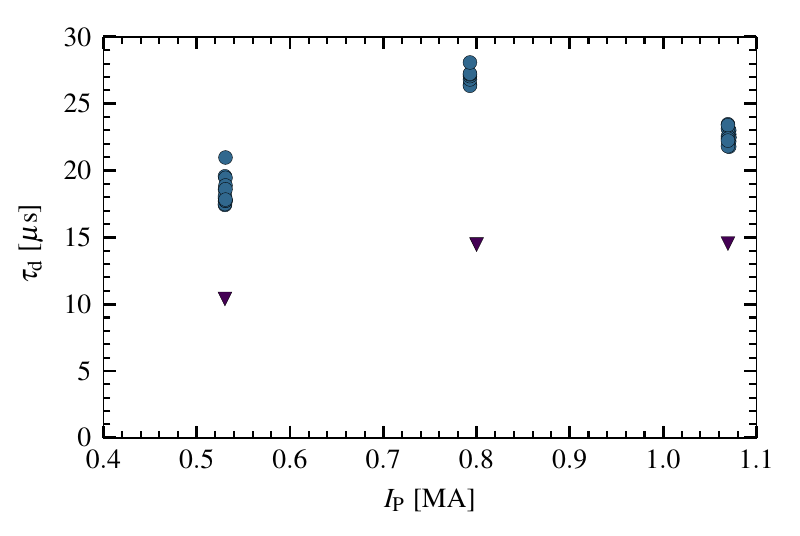}\label{fig:td-vs-current}}\\
        \subfigure[]{\includegraphics[width=0.48\textwidth]{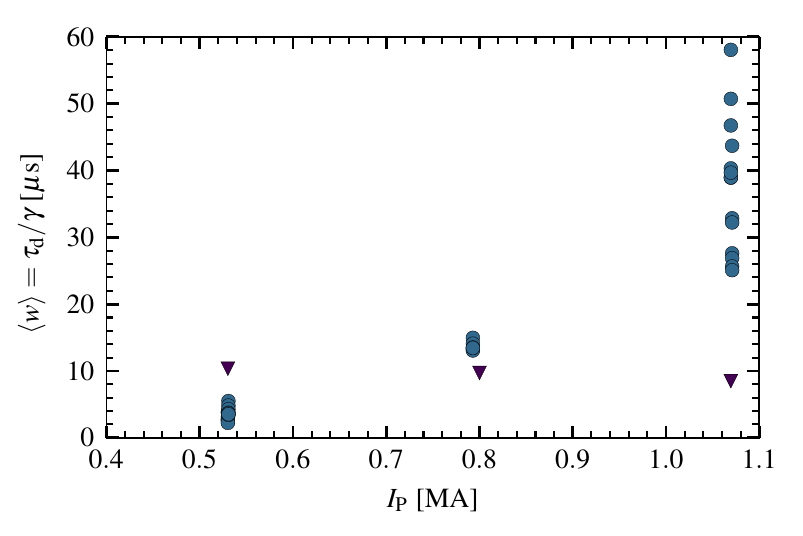}\label{fig:mean-wait-stats-vs-current}}
        \subfigure[]{\includegraphics[width=0.48\textwidth]{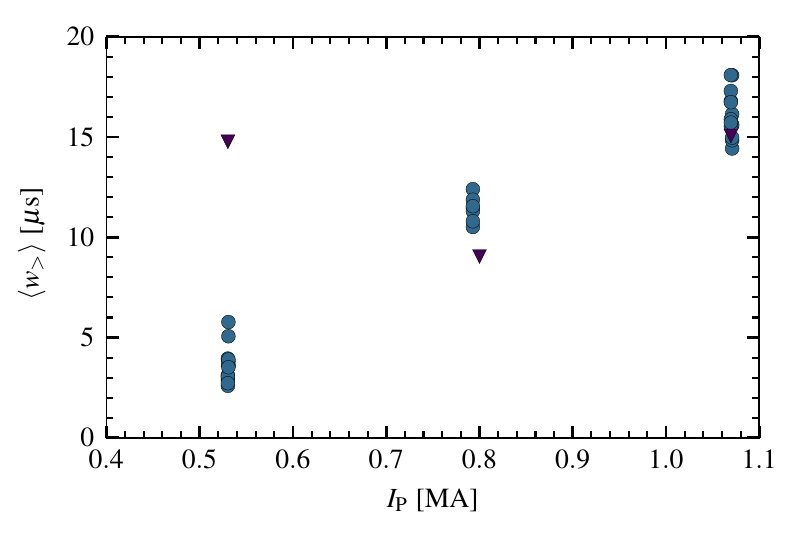}\label{fig:mean-wait-vs-current}}
    \caption{The stochastic model parameters plotted against plasma current. (a) Intermittency parameter, (b) the duration time, (c) mean waiting time from $\taud/\gamma$ and (d) estimated mean waiting time from the deconvolution, all as a function of plasma current showing estimated parameters from the GPI (blue circles) and the MLP $\Isat$ data (purple inverted triangles).}
    \label{fig:fpp-vs-current}
\end{figure}

The parameters of the FPP model are plotted against $\Ip$ in figure \ref{fig:fpp-vs-current} for both the GPI and the scanning MLP data. The means of the estimated amplitudes are not plotted because of the short time series of the scanning MLP data and questionable absolute values of the GPI light intensity measurements. For the GPI results shown here, we present all the diode views considered at time-averaged $\rho$ values ranging between $1.6-2.0\,\text{cm}$. The GPI shows that as $\Ip$ is changed, the parameters of the FPP model estimated from the time series are impacted. In particular, we observe stronger intermittency in the time series with increasing $\Ip$ shown in figure \ref{fig:gamma-vs-current}. In contrast, the MLP $\Isat$ measurements do not show such trends between the $\gamma$ and $\Ip$. This is consolidated by the profiles of the relative fluctuation levels showing indifference to the change in $\Ip$, as shown in figure \ref{fig:asp-mlp-current-scan-profiles}. The contradiction between these two diagnostics highlights the differences between them for fluctuation analysis, where the MLP is more of a localized measurement of plasma parameters compared to the GPI which is strongly impacted by spatial averaging. 

In figure \ref{fig:td-vs-current} for both diagnostics, $\taud$ seems independent of $\Ip$ where the MLP duration times are shorter compared to the GPI, which we also observe in the line-averaged density scan in figure \ref{fig:td-vs-fgw}. The GPI $\taud$ estimates are consistently a factor of approximately two larger than the MLP $\taud$ estimates, once again due to spatial averaging. These results imply that the plasma current does not impact the filament duration times estimated from the GPI and MLP.

The mean values of the estimated waiting time for both diagnostics are presented in figure \ref{fig:mean-wait-vs-current}. The GPI shows an increase in the mean estimated waiting time with $\Ip$, further suggesting that fluctuations become more intermittent with $\Ip$. This is consistent with decreasing intermittency (fluctuations becoming strongly intermittent) as seen in figure \ref{fig:gamma-vs-current}. However, for the MLP, no such trends are observed and the values for $\avg{w_>}$ seem to be higher compared to the GPI. Compared to the mean waiting time estimated from $\avg{w} = \taud/\gamma$ as shown in figure \ref{fig:mean-wait-stats-vs-current}, the MLP $\avg{w}$ also shows no change with $\Ip$. The GPI shows an increasing trend when it comes to the mean waiting times. There is a notable spread for $\avg{w}$ in the GPI data for $\Ip = 1.07\,\text{MA}$. In any case, the GPI results with $\avg{w} < 30$ $\mu$s show agreement within a factor of two with $\avg{w_>}$ in figure \ref{fig:mean-wait-vs-current}. 

\section{Issues with fit results}\label{appendix:fit-issues}

\begin{figure*}
    \centering
    \subfigure[]{\includegraphics[width=0.48\textwidth]{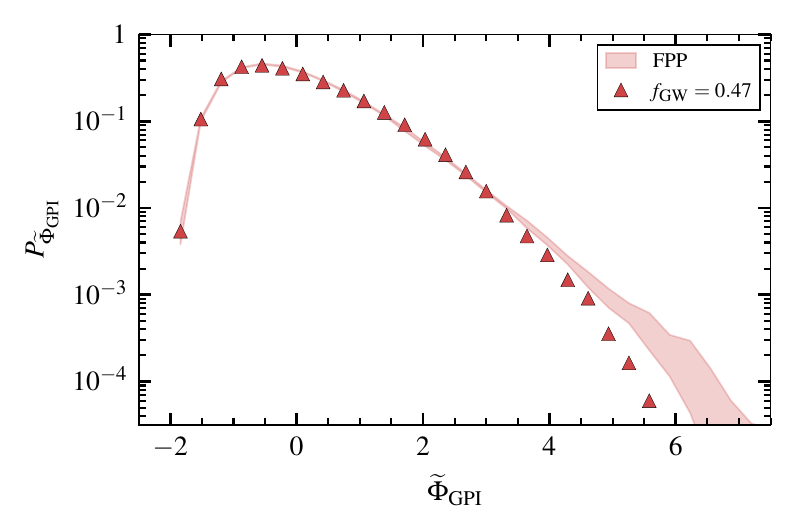}
    \label{fig:gpi-pdf-sig-18}}
    \subfigure[]{\includegraphics[width=0.48\textwidth]{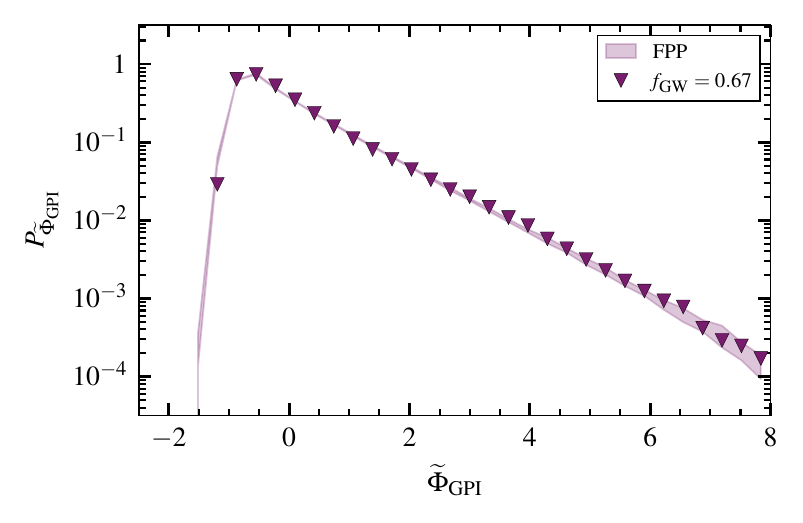}\label{fig:gpi-pdf-sig-22}}\\
    \caption{Histograms of the normalized GPI measurements for (a) $\fg$ = 0.47 and (b) $\fg$ = 0.67. The markers show the measurement data whereas the shaded regions show the minimum and maximum of the FPP histograms.}
    \label{fig:gpi-pdf-sig-fpp}
\end{figure*}
As an example, we will thoroughly describe the data analysis performed on GPI light intensity measurements from plasma discharges with Greenwald fractions of $\fg$ = 0.47 and $\fg$ = 0.67. We use the LMFIT module, which is a non-linear least squares optimization method in order to estimate the parameters of the stochastic model \cite{newville2016lmfit}. These will be then compared to synthetic realizations from the stochastic model, where we randomly generate around 10 realizations using the estimated $\gamma$ and $\eps$ from the ECF and the estimated $\taud$ and $\lambda$ from the PSD. These realizations assume observational noise.

In figure \ref{fig:gpi-pdf-sig-fpp} we show the histogram of the normalized measurements from the GPI for one pixel, for simplicity. We estimate $\gamma$ and $\eps$ from the empirical characteristic function, as it does not rely on the binning procedure. We will address the implications of this compared to the PDF estimate of the intermittency parameter later on. For the $\fg$ = 0.47 plasma discharge in figure \ref{fig:gpi-pdf-sig-18}, the PDF of the measurement data lacks an elevated tail. This makes it challenging to perform a parameter estimation. The FPP histograms in figure \ref{fig:gpi-pdf-sig-18} agree with the measurement data for approximately two decades in probability, but show an elevated tail in the PDF. However, for the Greenwald fraction-case of $\fg$ = 0.67 in figure \ref{fig:gpi-pdf-sig-22}, we see an elevated tail of the distribution and the agreement of the FPP histograms aligns well with the histogram of the measurement data. This demonstrates that the onus is on the quality of these fits, which can impact later results -- in particular for the intermediate Greenwald fraction discharges where these non-elevated tails are present, despite performing the analysis on long time-series measurements. 
\begin{figure*}
    \centering
    \subfigure[]{\includegraphics[width=0.48\textwidth]{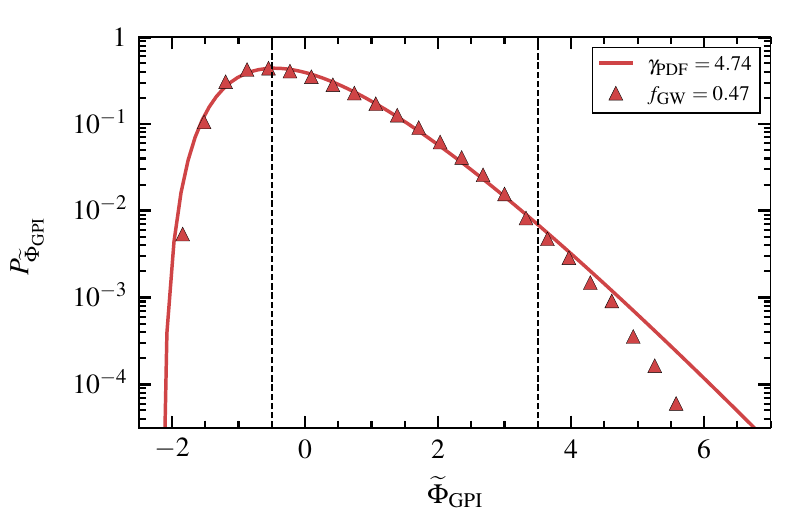}\label{fig:gpi-pdf-sig-18-small-range}}
    \subfigure[]{\includegraphics[width=0.48\textwidth]{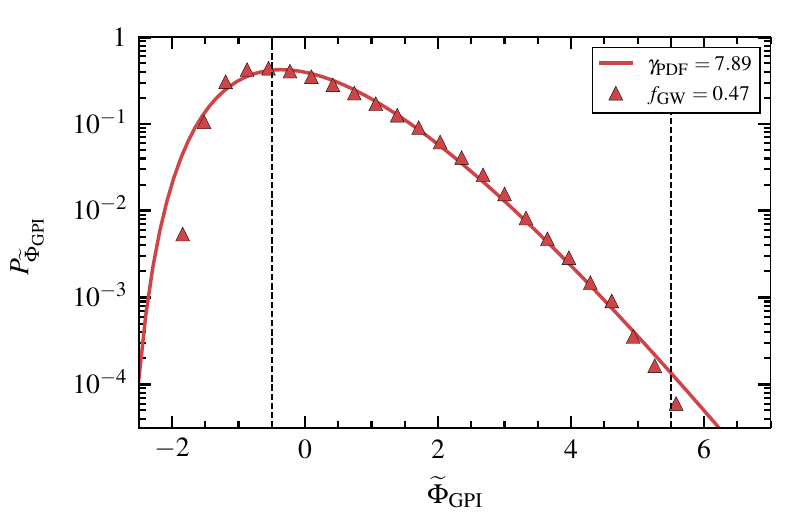}\label{fig:gpi-pdf-sig-18-large-range}}
    \caption{Histogram of the normalized GPI measurements for $\fg$ = 0.47 showing the fits based on the intervals chosen. Smaller fitting range is shown in (a) whereas in (b) a larger fitting range is used. The solid lines show the fits, the black dashed lines show the interval shown for performing the fit and the triangular markers represent the measurement data.}
    \label{fig:gpi-pdf-sig-fits-18}
\end{figure*}

Additionally, placing the weight on the tail impacts the way the fit agrees with the peak of the distribution. Figure \ref{fig:gpi-pdf-sig-fits-18} demonstrates this issue. Once again, we focus on the same pixel as before. Other pixels considered in the analysis later also show non-elevated tails in the PDF of the normalized signal. Here we use the analytical expression of the PDF from Equation (A9) in \cite{theodorsen-2017-ps}. Considering the fit in figure \ref{fig:gpi-pdf-sig-18-small-range}, where some of the tail is not included in the fitting procedure. Figure \ref{fig:gpi-pdf-sig-18-large-range} considers a wide range of values to fit the histogram. It is noticable that the peak is not well described by the fit using a large range, and hence the intermittency parameter being larger. We present these intermittency parameters and noise parameters from the PDF fits as $\gamma_\mathrm{PDF}$ and $\eps_\mathrm{PDF}$, respectively. For this reason, we chose to use the ECF to perform a parameter estimation for the intermittency and noise parameters in the rest of the study.

\begin{figure*}[ht!]
    \centering
    \subfigure[]{\includegraphics[width=0.48\textwidth]{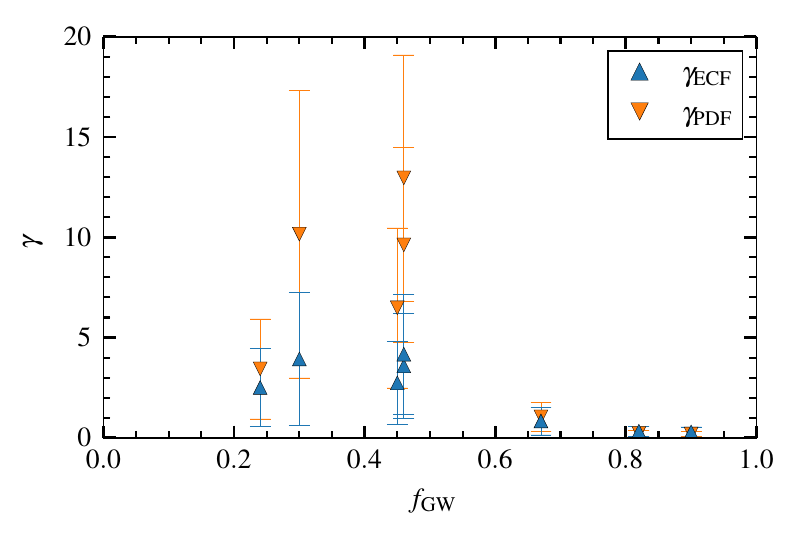}\label{ref:gpi-diff-gamma}}
    \subfigure[]{\includegraphics[width=0.48\textwidth]{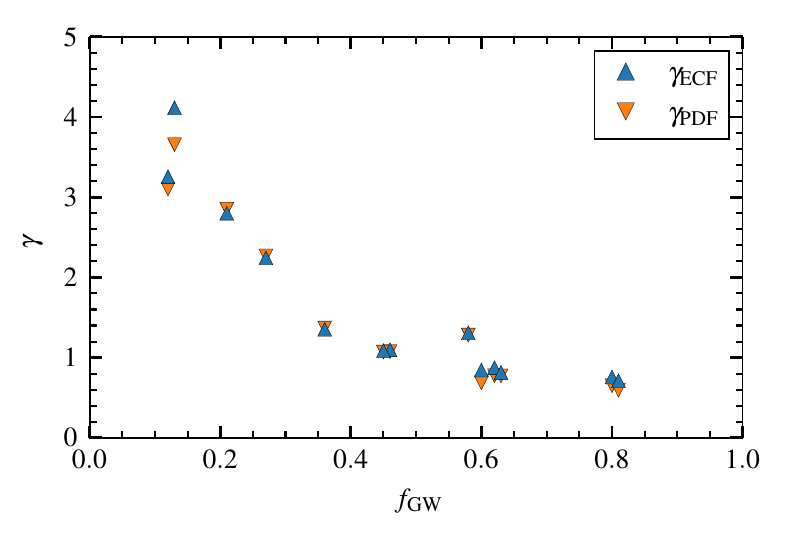}\label{ref:mlp-diff-gamma}}
    \caption{Comparison of the intermittency parameter estimate using the empirical characteristic function $\gamma_\mathrm{ECF}$ (blue triangles) and the PDF expression of the normalized time series $\gamma_\mathrm{PDF}$ (orange inverted triangles), applied to (a) the GPI measurements, where $\gamma_\mathrm{PDF}$ is estimated through a tail fit and (b) the MLP $\Isat$ data where $\gamma_\mathrm{PDF}$ is estimated with a similar fit range to figure \ref{fig:gpi-pdf-sig-18-small-range}. The markers represent the average $\gamma$ of all the APD views considered, whereas the errors bars are the minimum and maximum values of the estimated $\gamma$.}
    \label{fig:density-scan-diff-gamma-errors-vs-fgw}
\end{figure*}

Figure \ref{fig:density-scan-diff-gamma-errors-vs-fgw} shows the estimation of the intermittency parameter with the ECF compared to the analytical expression of the PDF for the GPI measurements (only to the tail as seen in figure \ref{fig:gpi-pdf-sig-18-large-range}) and the MLP $\Isat$ measurements (fitting ranges similar to figure \ref{fig:gpi-pdf-sig-18-small-range}). The expression of the PDF can be found in equation A9 of \cite{theodorsen-2017-ps}. For GPI measurements, a similar spread was also observed in the estimated intermittency parameter using the PDF as shown in figure \ref{fig:density-scan-diff-gamma-errors-vs-fgw}, where these values are larger compared to the ECF estimates. The intermittency parameter estimated from the MLP $\Isat$ measurements shows better agreement between both methods. Some variation can be observed between the two at lower densities. Therefore, to treat the GPI and MLP data in the same way, the ECF is used as opposed to the PDF since the intermittency parameter estimate is not sensitive to the number of bins. Overall, our impression of this study suggests that the estimation of parameters on the GPI data for $0.3 \leq \fg \leq 0.47$ is questionable due to the non-existent tails seen from the histograms. Otherwise, the intermitency parameters between the GPI and the MLP $\Isat$ align well for $\fg > 0.6$.

 \begin{figure}
    \centering
    \subfigure[]{\includegraphics[width=0.48\textwidth]{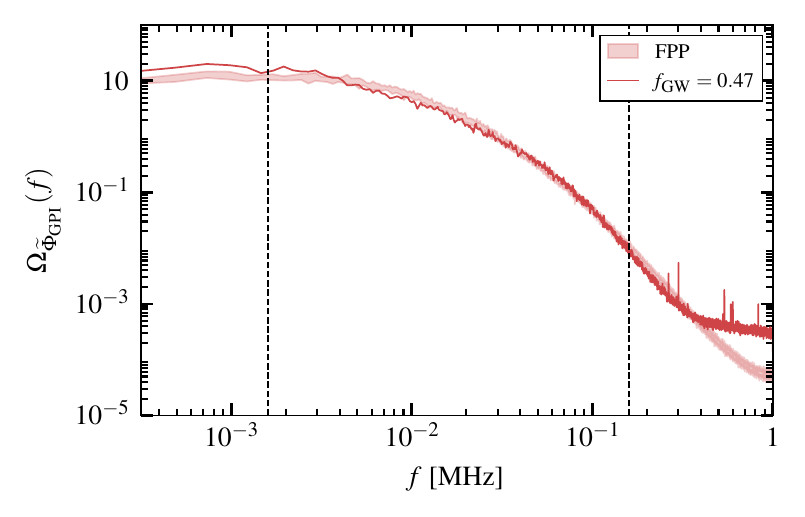}\label{fig:gpi-psd-sig-18}}
    \subfigure[]{\includegraphics[width=0.48\textwidth]{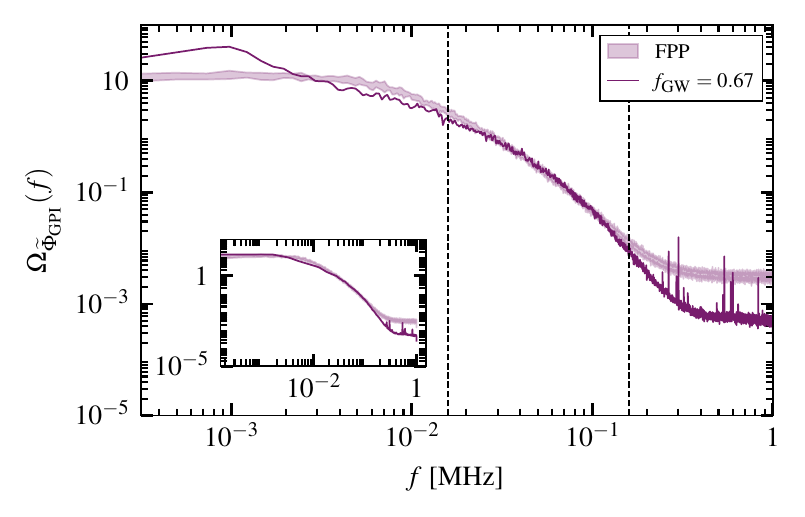}\label{fig:gpi-psd-sig-22}}
    \caption{PSDs of the normalized GPI measurements for (a) $\fg = 0.47$ and (b) $\fg = 0.67$, represented by the solid darker lines. The shaded regions are the minimum and maximum of the PSD of the FPP realizations. The dashed lines show the range considered for performing the fits.}
    \label{fig:gpi-psd-sig-fpp}
\end{figure}

We show the frequency power spectral densities (darker solid lines) of the normalized GPI measurements in figure \ref{fig:gpi-psd-sig-fpp} as well as the agreement with the parameters estimated from these FPP realizations, shown by the shaded regions. Welch's method was used to produce these spectra, which can be readily accessed through Python's SciPy package \cite{welch1967use}. The number of samples per segment (nperseg) for Welch's method on all discharges using the GPI was kept the same, which was nperseg = 8196. In figure \ref{fig:gpi-psd-sig-18}, the PSD of the measurement data from the $\fg = 0.47$-case agrees well with the realizations of the stochastic model. We see a clear bump in the power spectra in figure \ref{fig:gpi-psd-sig-22} for the $\fg = 0.67$ plasma discharge, which is located at $\sim1\,\text{kHz}$. This bump becomes prominent as the core density increases. The slope of the spectra is captured well by the fitting function, but struggles for the low-frequency part. In order to estimate $\taud$ without the influence of the bump, we lower nperseg to a point where the bump has been averaged out where this is shown in the inset of figure \ref{fig:gpi-psd-sig-22}. Indeed, it can be seen that the bump in the flat part of the spectra is substantially reduced and agrees well with the PSDs of the FPP realizations. We do not use such a low nperseg in the actual analysis since much of the data is smoothed away. Furthermore, a shorter fit range is used for $\fg = 0.67$ since the fit will overestimate $\taud$. We have discussed the consequences of overestimating $\taud$ for the pulse function in the deconvolution algorithm in reference \cite{ahmed-2022}.

\begin{figure}
    \centering
    \subfigure[]{\includegraphics[width=0.48\textwidth]{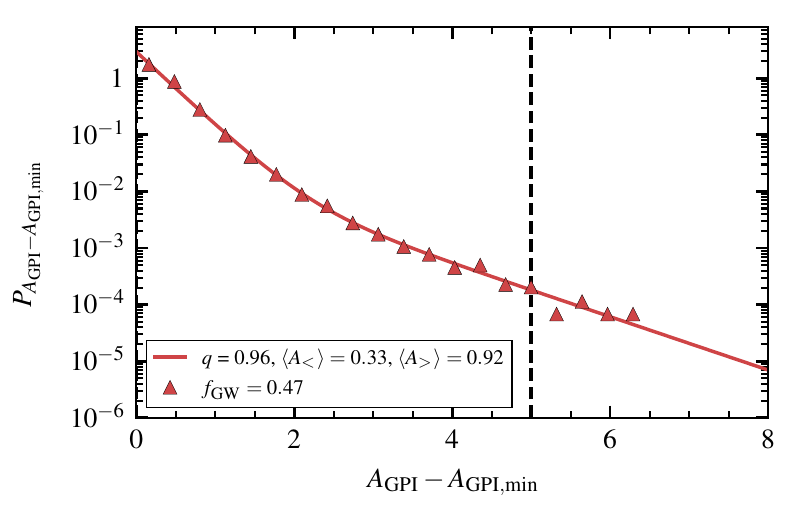}}\label{fig:gpi-bi-exp-amp-dist-18}
    \subfigure[]{\includegraphics[width=0.48\textwidth]{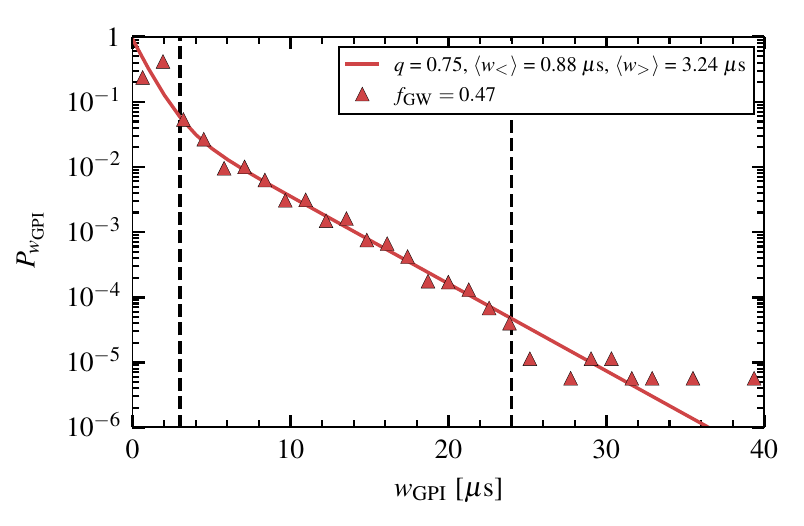}\label{fig:gpi-bi-exp-wait-dist-18}}
    \caption{Bi-exponential fits to (a) the estimated amplitude distribution and (b) waiting time distribution of the GPI measurements for $\fg=0.47$. The markers represents the deconvolved data. The black-dashed lines show the ranges taken into consideration for the fitting prodecude. The solid line represents the fits with the bi-exponential expression found in \eqref{eq:bi-exp-amp}.}
    \label{fig:gpi-bi-exp-amp-wait-dist-18}
\end{figure}

In addition, we present in the figure \ref{fig:gpi-bi-exp-amp-wait-dist-18}, the bi-exponential fits to the amplitude distribution and the waiting time distribution from the RL deconvolution on the GPI measurements for $\fg=0.47$ only. The bi-exponential fits shown by the solid lines were made to non-rescaled distributions in order to extract the mean of the estimated amplitude and estimated waiting time. As explained above, we are interested in large-amplitude events, and therefore we consider the mean of the estimated large amplitudes $\avg{A_{>}}$. Bear in mind that we do not trust these values to be the true mean amplitudes of the GPI measurements, since the measurements are not a proxy for plasma parameters. We see from figure \ref{fig:gpi-bi-exp-amp-dist-18}, the bi-exponential fit to the non-rescaled amplitude distribution. We present the parameters estimated for the partition between small and large amplitudes $q = 0.96$, the mean of the small amplitudes $\avg{A_<} = 0.33$ as well as $\avg{A_>} =0.92$. $q$ changes little with Greenwald fraction, therefore no results are shown for this. In figure \ref{fig:gpi-bi-exp-amp-wait-dist-18}, we present the bi-exponential fit to the estimated waiting time distribution using the RL deconvolution algorithm on the GPI measurement. The mean of the short waiting times was estimated to be $\avg{w_<} = 0.88$ $\mu$s and the mean of the long waiting times was found to be  $\avg{w_>} = 3.24$ $\mu$s.

\section*{References}
\bibliography{sources}
\bibliographystyle{iopart-num}
\end{document}